\documentclass{article}

\usepackage{microtype}
\usepackage{graphicx}
\usepackage{subfigure}
\usepackage{todonotes}
\usepackage{epstopdf}
\usepackage{amssymb}
\usepackage{booktabs} 
\usepackage{listings}

\usepackage{hyperref}

\usepackage[accepted]{sysml2019}

\sysmltitlerunning{Priority-based Parameter Propagation for Distributed DNN Training}
\newcommand{\name}{\textit{P3 }}
\newcommand{\named}{\textit{P3}}

\usepackage[firstpage]{draftwatermark}
\SetWatermarkText{
	\hspace*{3.7in}
	\raisebox{10in}{
		\includegraphics[height=0.9in]{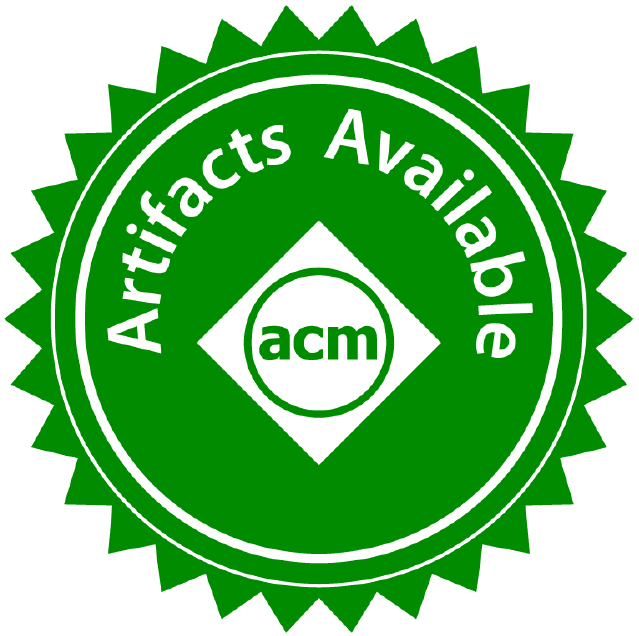}
		\includegraphics[height=0.9in]{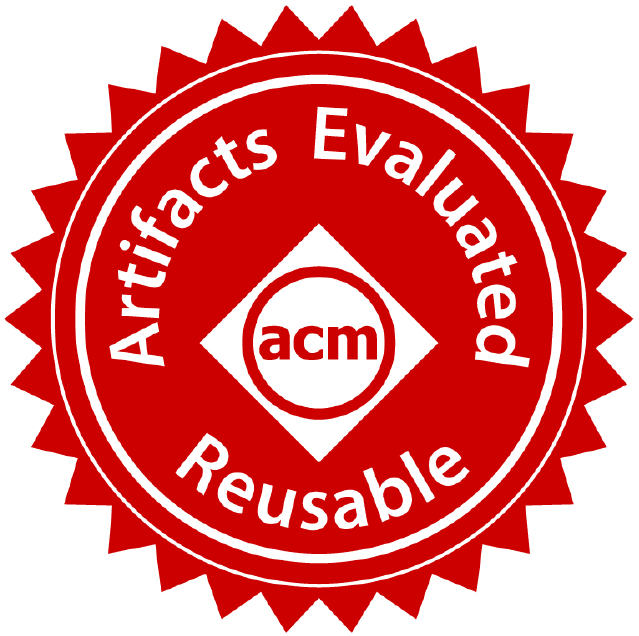}
	}
}

\begin{document}

\SetWatermarkAngle{0}

\twocolumn[
\sysmltitle{Priority-based Parameter Propagation\\for\\ Distributed DNN Training}




\begin{sysmlauthorlist}
\sysmlauthor{Anand Jayarajan}{ubc,vi}
\sysmlauthor{Jinliang Wei}{cmu}
\sysmlauthor{Garth Gibson}{vi,cmu}
\sysmlauthor{Alexandra Fedorova}{ubc}
\sysmlauthor{Gennady Pekhimenko}{uoft}
\end{sysmlauthorlist}

\sysmlaffiliation{ubc}{University of British Columbia, Vancouver, Canada}
\sysmlaffiliation{vi}{Vector Institute, Toronto, Canada}
\sysmlaffiliation{cmu}{Carnegie Mellon University, Pittsburgh, USA}
\sysmlaffiliation{uoft}{University of Toronto, Toronto, Canada}

\sysmlcorrespondingauthor{Anand Jayarajan}{anand.indukala@gmail.com}

\sysmlkeywords{Machine Learning, SysML}

\vskip 0.3in

\begin{abstract}
Data parallel training is widely used for scaling distributed deep neural network (DNN) training. However, the performance benefits are often limited by the communication-heavy parameter synchronization step. In this paper, we take advantage of the domain specific knowledge of DNN training and overlap parameter synchronization with computation in order to improve the training performance. We make two key observations: $(1)$ the optimal data representation granularity for the communication may differ from that used by the underlying DNN model implementation and $(2)$ different parameters can afford different synchronization delays. Based on these observations, we propose a new synchronization mechanism called \textit{Priority-based Parameter Propagation} (\named). \name synchronizes parameters at a finer granularity and schedules data transmission in such a way that the training process incurs minimal communication delay. We show that \name can improve the training throughput of ResNet-50, Sockeye and VGG-19 by as much as $25\%$, $38\%$ and $66\%$ respectively on clusters with realistic network bandwidth.
\end{abstract}
]



\printAffiliationsAndNotice{} 

\section{Introduction}\label{sec:intro}
Training DNN is notoriously time consuming because of the high degree of computational complexity involved in tuning billions of parameters and processing large amount data \cite{mlperf}. Data parallel distribution with synchronous stochastic gradient descent (SGD) is a popular method for accelerating the training by parallelizing the process over a cluster of machines \cite{ssgd}.

In this paradigm, the input data set is sharded among the worker machines and they train a shared DNN model iteratively by independently computing the parameter updates and synchronizing them at the end of every iteration. A single iteration on a worker machine involves three main steps: $(1)$ \textit{forward propagation} step for calculating the value of a loss function on a subset of local data shard, $(2)$ a subsequent \textit{backward propagation} step for computing the gradients for every model parameter based on the computed loss, and $(3)$ a \textit{parameter synchronization} step for aggregating local gradients from all the worker machines and updating the parameters using the SGD algorithm \cite{sgd}.

On every iteration, each worker machine generates and synchronizes hundreds of megabytes of gradient values \cite{nevol}. This often makes data parallel training a communication-bound workload \cite{tbd}. Handling such huge volume of traffic require high bandwidth networks like Gigabit Ethernet \cite{gigeth} or InfiniBand \cite{ib}. However, these technologies are yet to be adopted widely because of the high deployment cost. Moreover, this problem has been exacerbated with the emergence of faster hardware accelerators and larger DNNs as it leads to increase in data transmission rate and volume \cite{gpu-speed, moe}. Most of the major cloud providers and academic clusters are having trouble in catering to such high bandwidth demands \cite{phub}. In this work, we propose solutions to scale data parallel training under limited bandwidth conditions.

Gradient compression is one of the popular approaches aimed at reducing the communication overhead. Since gradient values are generally represented as floating point numbers, it is extremely challenging to get reasonable compression ratios from lossless compression techniques \cite{fpc}. Instead, recent studies propose lossy compression techniques like gradient quantization \cite{1bit,qsgd,tern} and sparse parameter synchronization \cite{gdrop,dgc}. These methods, however, risk affecting the final convergence accuracy of the model because of the information loss that comes with the value approximation and stale parameter updates \cite{loss}.

An orthogonal approach is to utilize the available network bandwidth more efficiently by leveraging domain specific opportunities in DNN training. The traffic generated by the training processes are generally bursty because of its iterative nature. A common practice used in some distributed machine learning (ML) frameworks is to attenuate these traffic bursts by overlapping communication with computation. Since the training computation is performed as a sequence of operations (\emph{layers}), during backward propagation, the gradients for the parameters associated with each layer are generated one after another. Frameworks such as TensorFlow \cite{tf}, MXNet \cite{mxnet} and Caffe2 \cite{caffe}, exploit this sequential layered structure to overlap parameter synchronization with backpropagation by issuing synchronization of each layer immediately after its gradients are computed.

\begin{figure}[h]
	\centering
	\includegraphics[width=70mm]{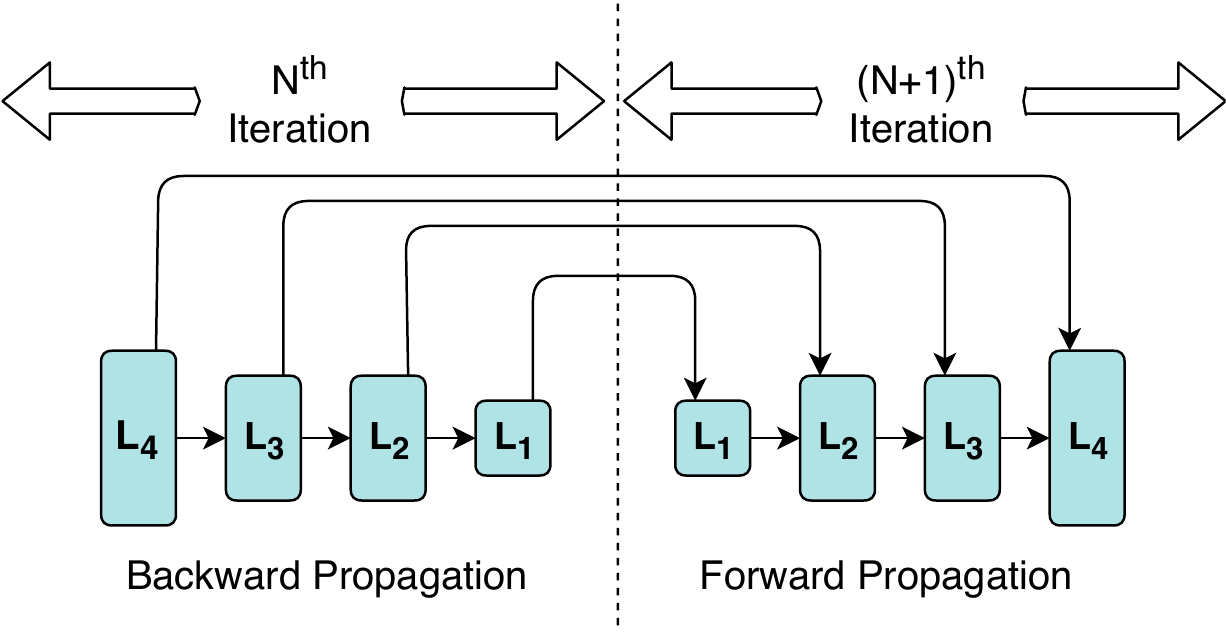}
	\label{fig:iter}
	\caption{Training iterations}
\end{figure}

In this work, we find new opportunities to better overlap communication and computation. Our first observation is that the domain specific knowledge of DNN training algorithm allows us to better schedule parameter synchronization not only based on when the gradients are generated, but also based on \emph{when the data is consumed}. During training, the gradients of the layers are generated from final to initial layers and subsequently consumed in the reverse order in the next iteration. Figure \ref{fig:iter} shows a snapshot of the training process containing the backward propagation of one iteration and the forward propagation of the next one. The temporal gap between gradients generated and consumed per layer are higher for final layers compared to the initial ones. Scheduling parameter synchronization using this information can help to overlap communication with both the forward and the backward propagation.

Secondly, the layer-wise granularity used by the underlying neural network implementation may not always be optimal for parameter synchronization. In our experiments, for certain heavy models (e.g., VGG \cite{vgg}, Sockeye \cite{sockeye}), parameter synchronization at a finer granularity improves the network utilization and reduces the communication delay.

Based on these observations, we propose a new synchronization mechanism called \textit{Priority-based Parameter Propagation} (\named).

\subsection{Our Approach}\label{sec:approach}
\name consists of two key components: $(1)$ \textit{Parameter Slicing}: \name splits the layers into smaller slices and synchronize them independently. $(2)$ \textit{Priority-based Update}: \name synchronizes the parameter slices based on their priority, where the priority of a slice is defined by when it is required again in the subsequent iteration. During backpropagation, \name always allocates network cycles to the highest priority slices in the queue, preempting synchronization of the slices from a previous lower priority layer if necessary.

\name offers following advantages over state-of-the-art parameter synchronization mechanisms \cite{poseidon, dgc}. $(1)$ \name can provide improved training performance under limited bandwidth conditions by better overlapping communication with computation and utilizing the available network bandwidth more efficiently. $(2)$ \name is model-agnostic, its implementation requires minimal programming effort, and all required changes are localized within the framework. $(3)$ \name always communicates full gradients and does not affect model convergence.

In summary, this paper makes the following contributions:
\vspace{-4pt}
\begin{itemize}
	\item{We show that parameter synchronization at layer-wise granularity can cause suboptimal resource utilization in heavy models (e.g., VGG, Sockeye). We also show that the parameter synchronization can be scheduled better to efficiently use the available network bandwidth by taking into account not only the information on when the gradients are generated, but also when they are consumed.}
	\vspace{-4pt}
	\item{We present a new parameter synchronization mechanism called \textit{Priority-based Parameter Propagation} (\named), which uses parameter slicing and priority-based updates to reduce communication overhead. We demonstrate that \name has better resiliency towards bandwidth limitations compared to other state-of-the-art synchronization mechanisms \cite{poseidon}.}
	\vspace{-4pt}
	\item{We implement and open source \named{\footnote{\href{https://github.com/anandj91/p3}{https://github.com/anandj91/p3}}} on MXNet \cite{mxnet}, a popular distributed ML framework, and evaluate its performance against standard MXNet implementation as the main baseline. We observe that, \name improves training performance of several state-of-the-art models like ResNet-50 \cite{resnet}, Sockeye \cite{sockeye} and VGG-19 \cite{vgg} by as much as $25\%$, $38\%$ and $66\%$ correspondingly.}
\end{itemize}

\section{Background}\label{sec:back}
\begin{figure}[h]
	\centering
	\includegraphics[width=60mm]{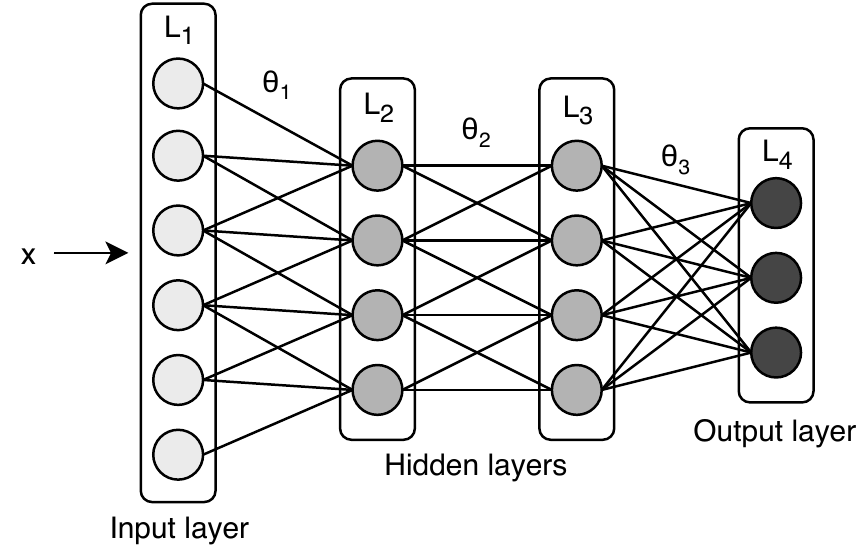}
	\caption{Deep neural network structure}
	\label{fig:dnn}
\end{figure}

The fundamental building blocks of DNNs are mathematical operations such as convolution, matrix multiplication, and activation functions. These operations perform certain transformations ($f_\theta(x)$) on an input vector ($x$) using the parameters ($\theta$) associated with it. A DNN is defined by a sequence of such operations (layers). In Figure \ref{fig:dnn}, the initial layer takes the application specific data samples as input and produces a prediction as an output vector at the final layer. The goal of the training algorithm is to find the parameter values which can make the most accurate predictions.

DNN training usually starts from a random parameter initialization and iterates by randomly sampling input vectors from the training dataset. On every iteration, the DNN computes the output vector on the inputs and calculates the error associated with the prediction (loss) by feeding the output vector to a loss function. This step is called a \emph{forward propagation}. After that, a \emph{backward propagation} step is performed that calculates the error contribution of each parameter by computing gradients of all the layers with respect to the loss. The backward propagation method for calculating gradients is based on the chain rule of derivatives and is therefore performed in the reverse order of forward propagation, i.e., gradients of the final layer is calculated first and moves backwards to the initial layers, hence the name backward propagation \cite{backprop}. Once the gradients are calculated, the parameters are updated using an optimization algorithm like Stochastic Gradient Descent (SGD) \cite{sgd}. The iterations are repeated several times over the training dataset until the model converges to an acceptable prediction accuracy.

Each iteration is highly compute-intensive which makes the training process very time consuming. The total training time can be dramatically reduced by distributing the workload into multiple machines by taking advantage of the data parallel nature of the SGD algorithm. Data parallel training \cite{datapar} involves multiple workers simultaneously training a shared DNN with the training dataset distributed equally among them. On each iteration, workers independently calculate the gradients locally for a common parameter value assignment but on different input data samples. Then the gradients are aggregated in a synchronous fashion for performing parameter updates. This method is called a \textit{synchronous SGD} algorithm \cite{ssgd}. 

Parameter server architecture \cite{pslite} is one of the most popular methods used in practice for parameter synchronization and it is widely supported in most of the distributed ML frameworks (e.g., MXNet \cite{mxnet}, TensorFlow \cite{tf}, Caffe2 \cite{caffe}). The parameter server keeps track of the up-to-date values of all the model parameters. Before every iteration, each worker machine reads the latest parameter values ($\theta$) from the parameter server and locally computes gradients for the inputs sampled from its data shard. The workers then send the local gradients ($\triangledown$) to the parameter server. The parameter server waits until it receives gradient updates from all worker machines, then aggregates the gradients together and updates the parameters for the next iteration.

\begin{figure}[h]
	\centering
	\includegraphics[width=50mm]{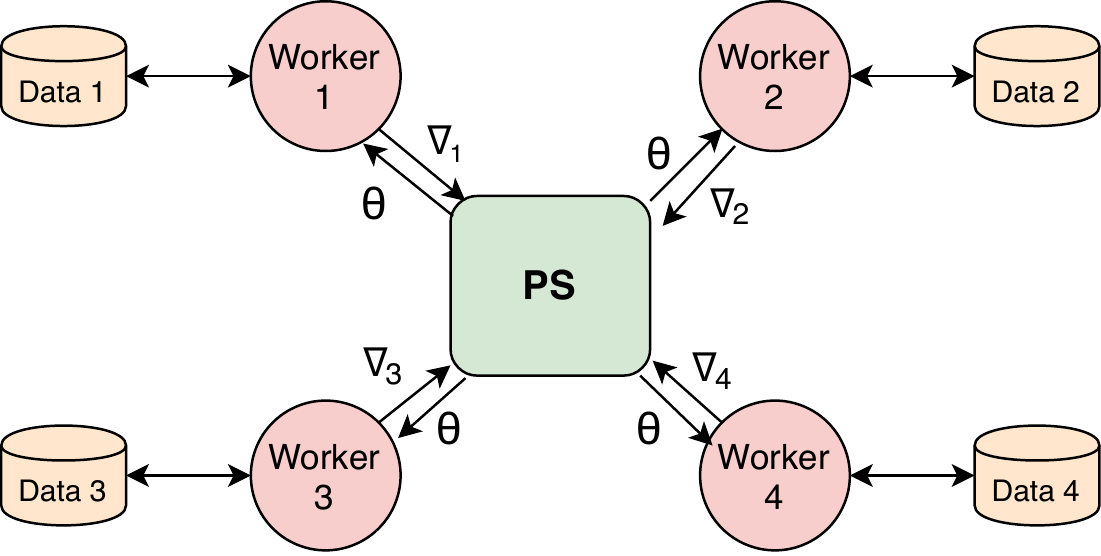}
	\caption{Parameter server architecture}
	\label{fig:ps}
\end{figure}

Figure \ref{fig:ps} shows parameter server-based data parallel training in a four-node cluster. The communication between worker machine and parameter server is usually over a network and often becomes the bottleneck in achieving linear scalability in data parallel training \cite{tbd,distbench}.

State-of-the-art ML framework MXNet is designed specifically for making data parallel training efficient and easy to execute. It comes with a built-in implementation of parameter server called \textit{KVStore}. In MXNet, worker machines send out gradients of a layer to the \textit{KVStore} as soon as they are calculated, and issue parameter pull requests once all the other workers have finished sending the gradients of that layer. This aggressive parameter synchronization makes data parallel training very efficient on MXNet.

TensorFlow, on the other hand, is designed as a more generic ML framework. Hence it does not have an explicit parameter server implementation. However, a parameter server can be implemented on top of the graph computation framework provided by TensorFlow. Since the parameter server is a part of the computation graph, the communication between the worker subgraph and parameter server subgraph is handled by the framework itself. TensorFlow automatically places \textit{Send} and \textit{Receive} operations on the edges of the computation graph that crosses the device boundaries. Similar to MXNet, the worker subgraph executes the send operation as soon as the gradients are computed. However, since every training iteration is a separate graph execution, the parameter pull request is not issued until start of the next iteration. This disconnection in sending gradients and receiving parameter updates could cause underutilization of bidirectional network bandwidth.

Despite small differences described above, we observe that state-of-the-art ML frameworks (e.g., MXNet, TensorFlow, Caffe2.) have two common characteristics. For performance reasons, the operations in the DNN implementation usually perform computations on large data representations. Therefore, the gradients for all the parameters within a layer are usually generated in a single shot. We observe that, because the gradients are generated at the layer level granularity, frameworks perform parameter synchronization at the same granularity as well. We also observe that since the DNN implementation is written as a dependency graph in these frameworks, the gradients of the layers are sent out to the parameter server over the network as soon as the backward propagation of that layer has completed. In this work, we address the limitations associated with these two characteristics of ML frameworks.

Apart from parameter server architecture, there are other mechanisms used for gradient aggregation. For example, there are many variations of MPI \textit{all\_reduce} operation specifically designed for ML workloads \cite{gossip,dense}. In this work, we implement \name over the parameter server architecture in MXNet. However, we believe, \name design principles (namely, parameter slicing and priority-based propagation) are general enough to be applied to any gradient aggregation methods.
\section{Limitations of parameter synchronization}\label{sec:lim}
Current parameter synchronization mechanisms have major limitations in effectively utilizing available network bandwidth due to two main reasons. The first one comes with the aggressive synchronization performed by the frameworks where the gradients of the layers are sent to the parameter server immediately after finishing the backward propagation of that layer. Since the backward propagation progresses from the final to the initial layer, the gradients are also generated and propagated in that order. However, the next forward propagation can only be started after receiving the updated parameters of the \emph{first layer}. We observe that, under limited bandwidth, gradient propagation of the final layers can induce queuing delay onto the gradient propagation of the initial layers and subsequently delay the next iteration. This prevents the communication from being overlapped with the forward propagation.

\begin{figure}[h]
	\centering
	\subfigure[Aggressive synchronization]{
		\label{fig:delay-bad}
		\includegraphics[width=60mm]{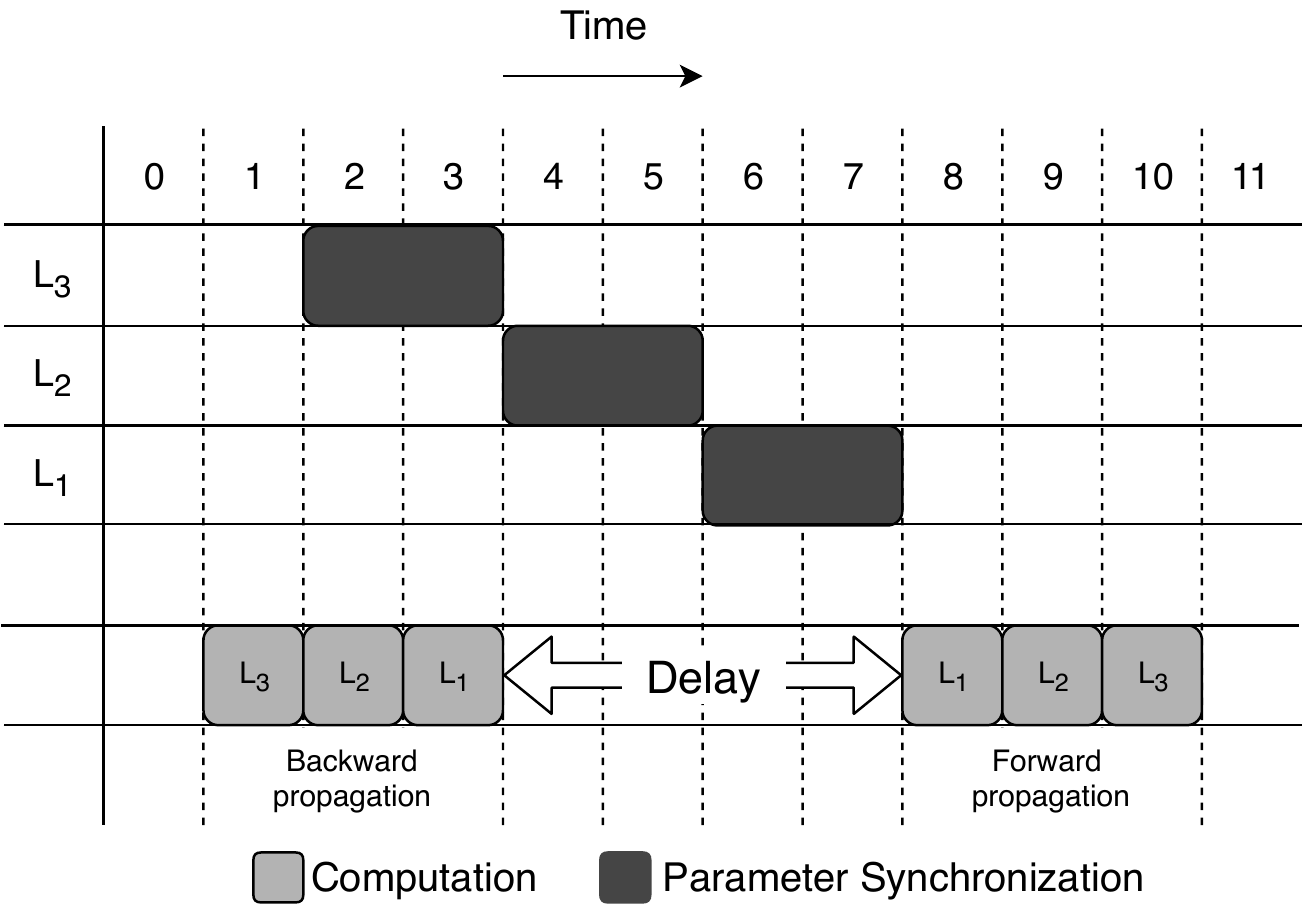}
	}
	\subfigure[Priority based synchronization]{
		\label{fig:delay-fix}
		\includegraphics[width=60mm]{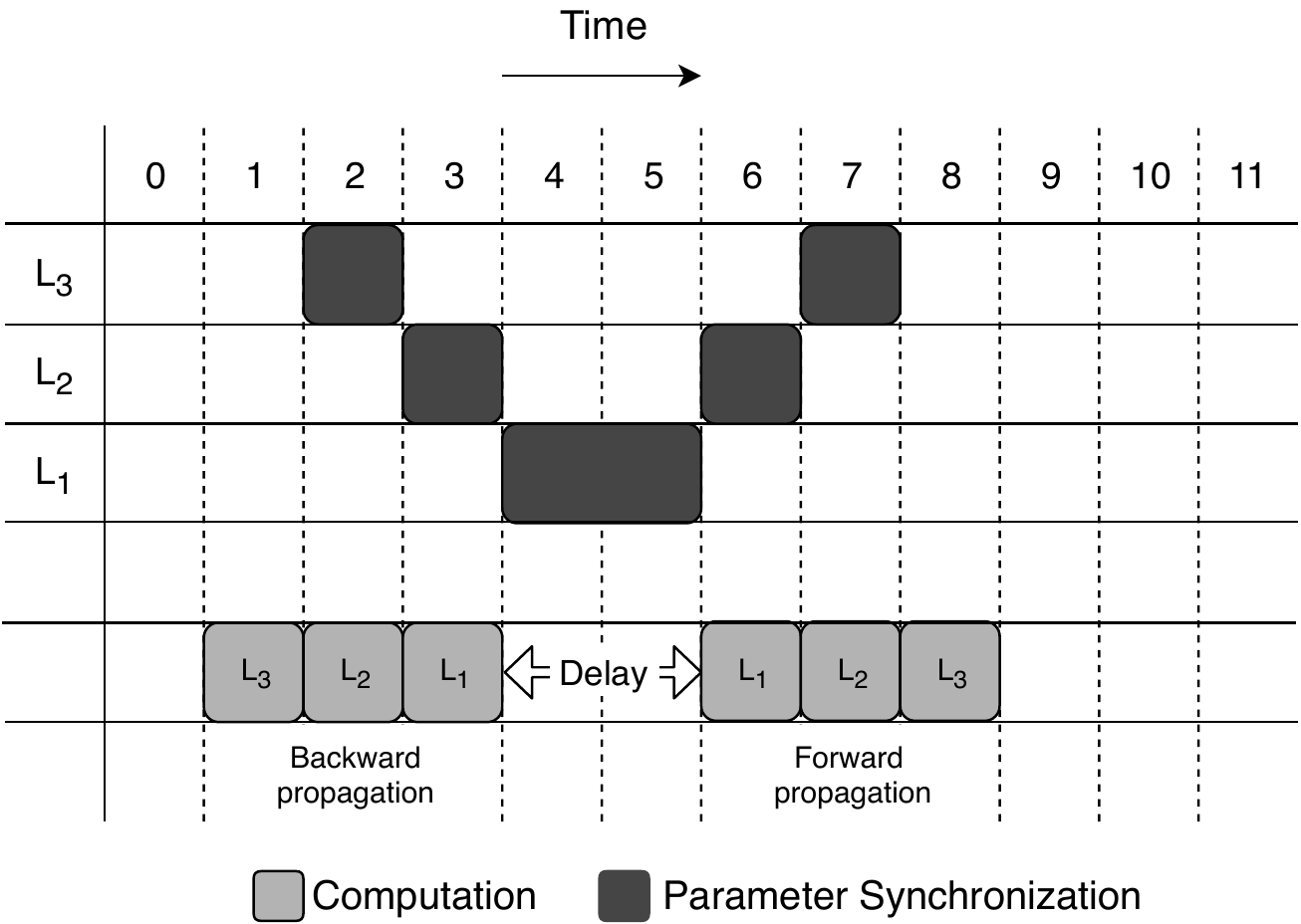}
	}
	\label{fig:delay}
	\caption{Parameter synchronization}
	\vspace{-5pt}
\end{figure}

Figure \ref{fig:delay-bad} shows the parameter synchronization of a 3-layer DNN, where the forward and backward propagation of each layer takes one time unit and parameter synchronization takes two time units. With aggressive synchronization, the total delay between the two iterations is twice the time taken for synchronizing the first layer because of the additional queuing delay induced by the previous layers. Moreover, during forward propagation the network stays totally idle.

\begin{figure*}[h]
	\centering
	\subfigure[ResNet-50]{
		\label{fig:dist-resnet}
		\includegraphics[width=50mm]{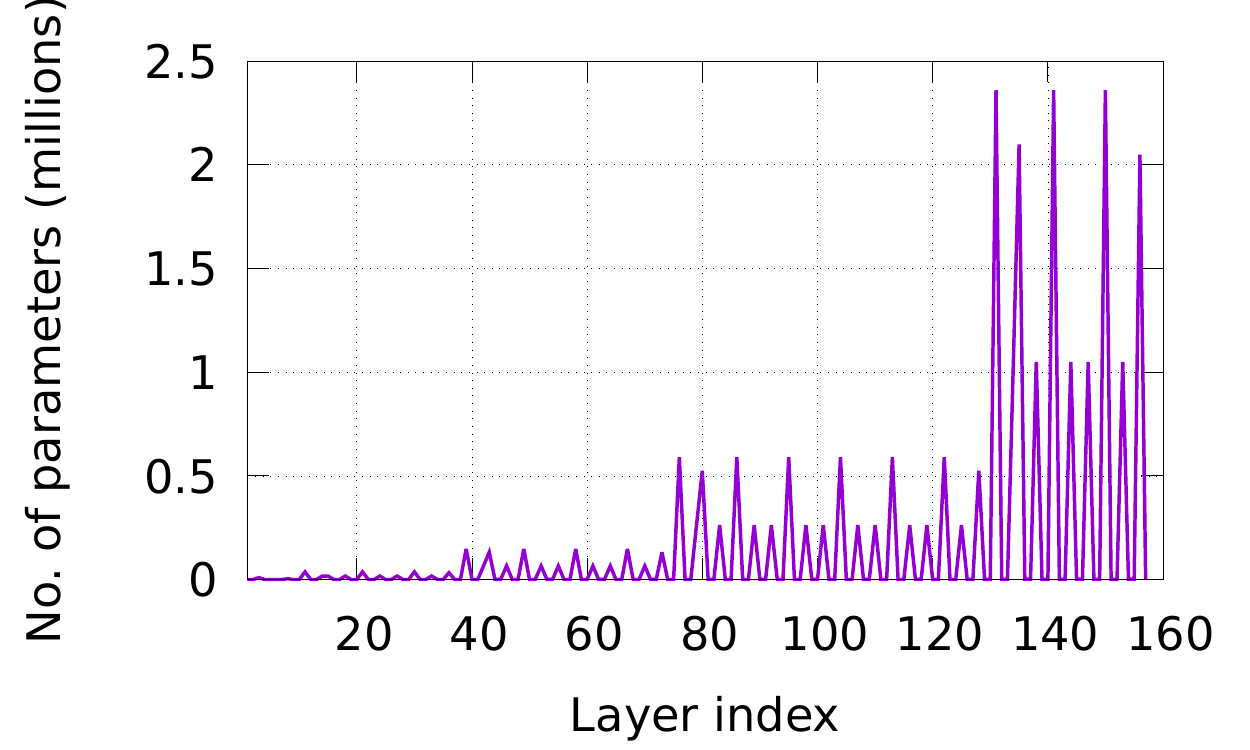}
	}
	\subfigure[VGG-19]{
		\label{fig:dist-vgg}
		\includegraphics[width=50mm]{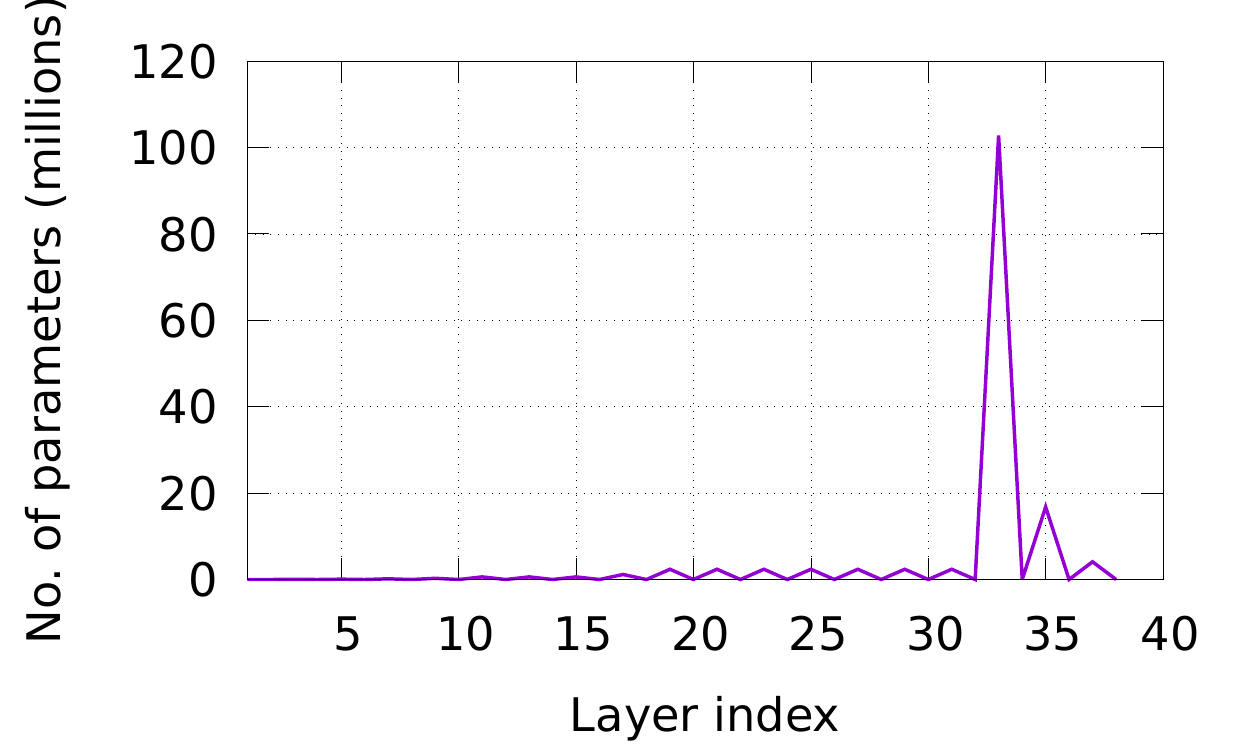}
	}
	\subfigure[Sockeye]{
		\label{fig:dist-sockeye}
		\includegraphics[width=50mm]{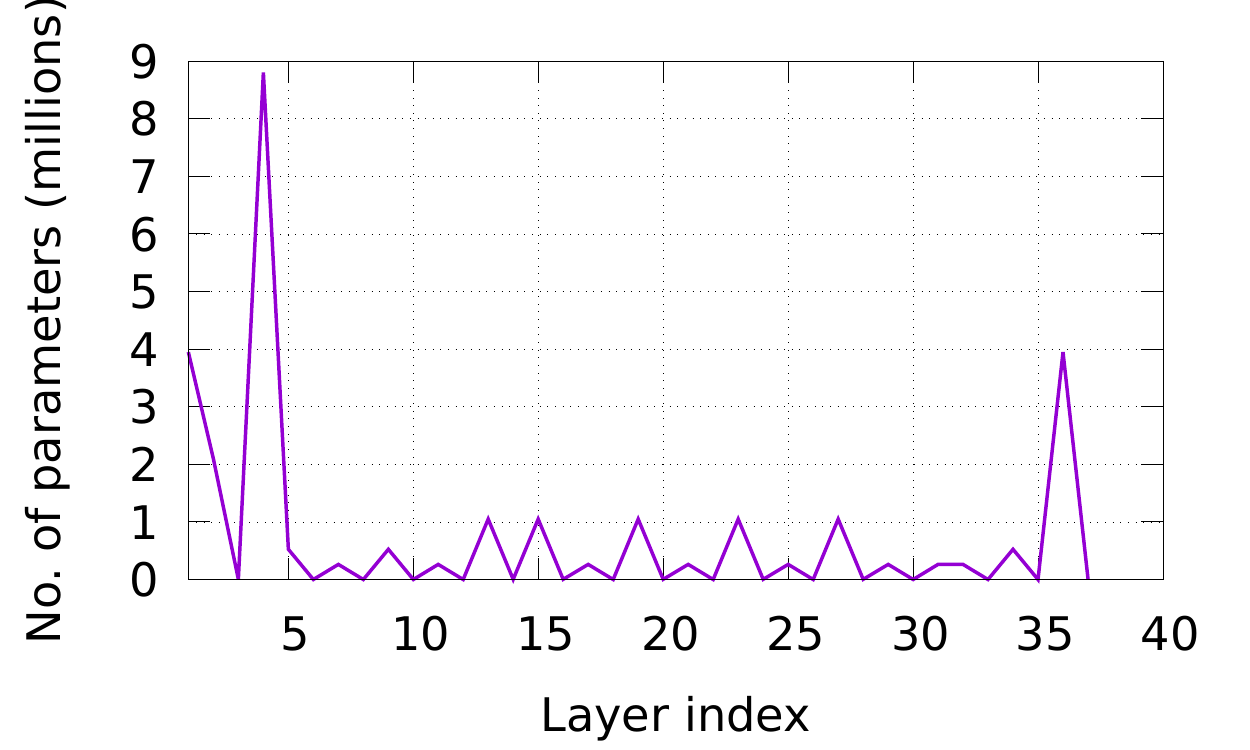}
	}
	\vspace{-5pt}
	\caption{Parameter distribution}
	\label{fig:dist}
	\vspace{-5pt}
\end{figure*}

This effect becomes even more noticeable when the communication time required for individual layers vary due to the presence of fully connected (FC) layers in the DNN, as the synchronization time needed for such dense layers is relatively higher. Figure \ref{fig:dist} shows the parameter distribution of two popular image classification models: ResNet-50, VGG-19, and a machine translation model: Sockeye. The skewed parameter size distribution is a general trend in image classification models where the final FC layers are usually heavier and can potentially induce higher queuing delays on to the lighter initial convolution layers.

The second limitation is due to the parameter synchronization being performed at layer-level granularity. The communication time of parameter synchronization consists of three major parts: $(1)$ \textit{gradient propagation} time for the worker machine to send the gradients to the parameter server, $(2)$ \textit{parameter update} time taken by the parameter server to aggregate the gradients and update the parameters, and $(3)$ \textit{parameter propagation} time taken by the parameter server to send the updated parameters back to worker machine(s). As we describe in Section \ref{sec:back}, current distributed ML frameworks overlap gradient propagation of one layer with the backward propagation of the next one. On top of this, at the parameter server side, the gradient propagation of a layer is overlapped with the parameter update of the previous layer. This type of communication-computation pipelining is effective only if the size of the layers are more or less uniform. Unfortunately, this is usually not the case. For example, Figure \ref{fig:dist-vgg} shows that VGG-19 contains a single FC layer which has $71.5\%$ of all the parameters in the entire network. We observe that the disproportionately heavy layers like this could severely affect the efficient utilization of network bidirectional bandwidth.

\begin{figure}[h]
	\centering
	\subfigure[Layer level granularity]{
		\label{fig:pipe-bad}
		\includegraphics[width=60mm]{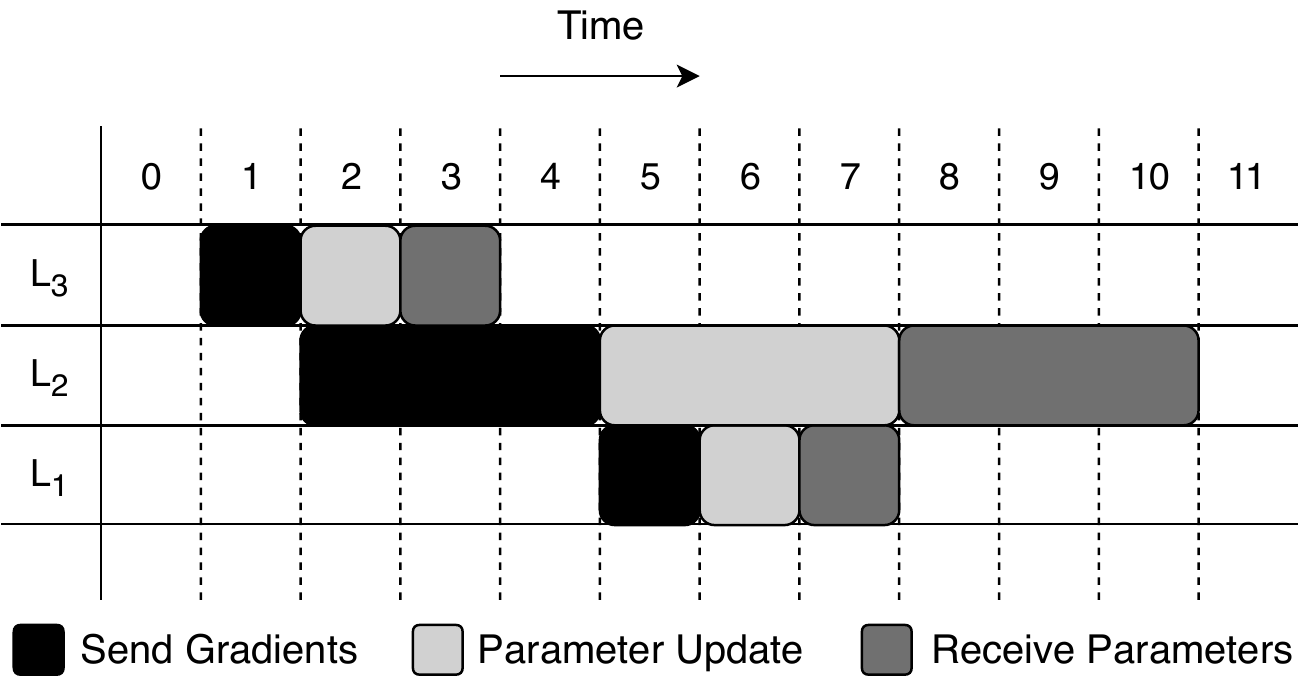}
	}
	\subfigure[Fine granularity]{
		\label{fig:pipe-fix}
		\includegraphics[width=60mm]{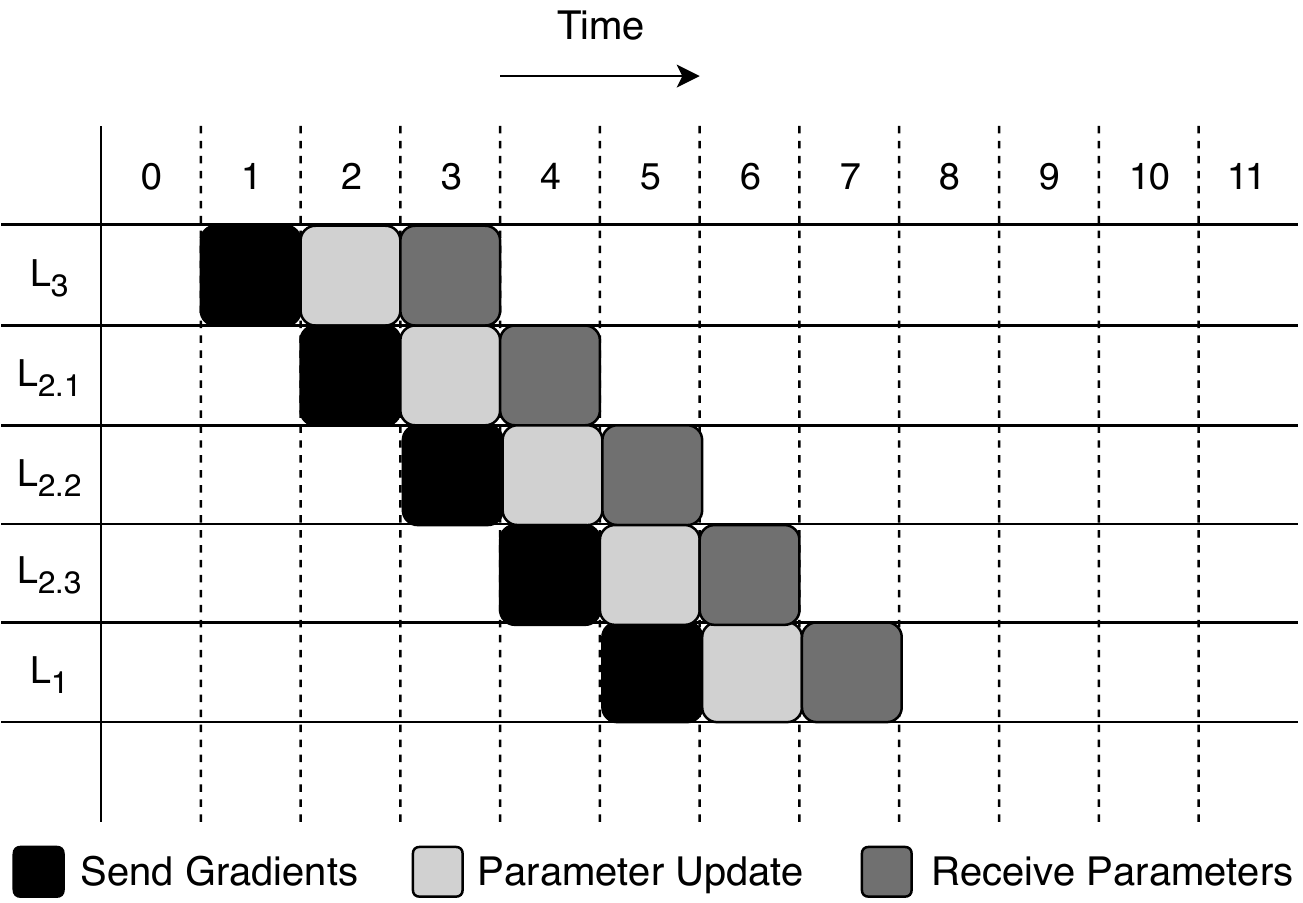}
	}
	\label{fig:pipe}
	\caption{Coarse and fine granularity}
\end{figure}

This effect is explained in Figure \ref{fig:pipe-bad} using the previous example of parameter synchronization of the 3-layered DNN. In this case, gradient propagation, parameter update, and parameter propagation of the second layer take thrice as much time as that of the first and third layers. Because of this imbalance, the communication delay in this model is mainly dominated by the second layer. The parameter synchronization of the first and the third layer can only be partially overlapped with the second layer. As seen in the example, this severely underutilizes the computing resources and bidirectional bandwidth by spending the last three time steps just for receiving parameter updates from the parameter server.

From the above observations we draw two major conclusions. First, the application domain-specific knowledge of DNNs can be utilized to schedule communication not only based on the data generation in the backward propagation, but also based on when the data consumption in the subsequent forward propagation. Scheduling parameter synchronization based on this information and sending the gradients conservatively could reduce the delay by better overlapping communication with both the forward and the backward propagation. Second, the optimal granularity required for parameter synchronization may differ from the one used for data representation by the underlying model implementation. Synchronizing parameters at a finer granularity can better utilize the available computing and networking resources as we empirically show in Section \ref{sec:netutil}.

\section{\named: Design and Implementation}\label{sec:prio} Based on the above
observations, we propose a new method for parameter synchronization called Priority-based Parameter Propagation (\named). As explained in Section \ref{sec:approach}, \name has two core components: $(1)$ \textit{parameter slicing}, and $(2)$ \textit{priority-based update}.

Once the gradients are computed, gradient aggregation and updates of each parameter can be performed independent to each other. We take advantage of this property of the SGD algorithm for parameter slicing optimization. \name splits the layers into smaller slices of parameters and each of these slices are then independently synchronized. In Figure \ref{fig:pipe-fix},  applying parameter slicing optimization on the second layer achieves better overlap between data transmission and parameter update. Moreover, the bidirectional bandwidth is completely utilized as the synchronization of slices are perfectly pipelined. In this example, parameter slicing reduces the communication cost by $30\%$.

After splitting the layers into smaller pieces, \name assigns priorities to every slice. The slices inherit priority values from their parent layers. We determine layers' priorities based on the order in which they are processed in the forward propagation. The first layer gets the highest priority and it decrements moving towards to the end, with final layer getting the lowest priority. During backward propagation, parameter synchronization of the slices are issued based on their priorities as illustrated in Figure \ref{fig:delay-fix}. In this example, with the priority-based update, the delay between the two iterations has been reduced by half and the communication is evenly overlapped with both the forward and the backward propagation.

We implemented \name by modifying the parameter server module in MXNet called KVStore. Below, we first explain how the baseline KVStore works, and then we describe our modifications to support \named.

\subsection{KVStore: Baseline system}\label{sec:kvs} KVStore is a wrapper implemented on top of the light-weight parameter server called \textit{ps-lite} \cite{pslite}. KVStore has two components: \emph{KVWorker}, which runs locally to the worker machine as part of the training process and a separate server process, called \emph{KVServer}. KVServer receives gradients from KVWorkers, aggregates them, and updates the parameters while ensuring data consistency. The parameters are stored as key-value pairs, where the key is the index of a layer and the value is an array of floating points each corresponding to the parameter values of that layer. For load balancing purposes, more than one KVServer can be used for the training with the parameters equally sharded between them. For better resource utilization, a common practice is to run one KVServer on every machine along with the worker process.

Before starting the training process, KVStore initializes and distributes the
parameters of all the layers among KVServers. KVStore follows a simple
heuristic for fair distribution of parameters. Layers with size smaller than a fixed threshold are assigned to a randomly chosen KVServer. Parameters of
larger layers are split equally among the KVServers. This is different from
parameter slicing used in \name (explained in Section \ref{sec:impl}). The
threshold is a configurable parameter and is set to $10^6$ parameters by
default.

KVServer exposes two interfaces to KVWorker for sending gradients and requesting updated parameters: a \emph{Push} request and a \emph{Pull} request. During training, MXNet issues a parameter synchronization request for a layer to the KVServer through the KVWorker as soon as the backpropagation of that layer has finished. KVWorker serializes (and fragments in case of large layers) the gradient matrix and issues a Push request to the corresponding KVServer(s). KVServer waits until it has received gradient updates from all the workers for that key-value pair. Once all the updates have been received, KVServer aggregates the gradients and updates the parameters. 

Once the parameters are updated, KVServer notifies all the workers. When KVWorker receives a notification, it immediately issues a Pull request to the KVServer(s) for the corresponding updated parameter values. KVServer then sends the latest parameter values in response, and KVWorker (reconstructs the parameter array for large layers and) updates the local parameter values for the next iteration. MXNet overlaps the parameter synchronization with backward propagation by asynchronously issuing Push requests for the layers whose gradients are ready to be propagated.

\subsection{\named: Implementation}\label{sec:impl} In order to implement
\named, we modify KVWorker and KVServer into {\named}Worker and {\named}Server.
On the worker side, when a parameter synchronization is issued, {\named}Worker
splits the gradient matrix of the layer based on a predefined size threshold
(explained in more details in Section \ref{sec:slice-size}). Unlike
in KVStore, this threshold defines the maximum granularity with which layers can be split. This is the parameter slicing part in \named. Each of these slices are assigned to {\named}Servers in a round-robin fashion.

The priority-based gradient propagation is implemented using a
producer-consumer mechanism communicating through a priority queue. After
parameter slicing, the producer part of {\named}Worker assigns priorities to the
individual slices and pushes them into the priority queue all at once. A
separate consumer thread in the {\named}Worker continuously polls the highest
priority slice from the queue and sends the slice to the {\named}Server through the network with its priority added to the packet header. The consumer thread uses blocking network calls. Hence the rate at which the priority queue polled is automatically adjusted based on the networking delay. This simple producer-consumer model makes sure that the network does not experience bursty traffic flow from the {\named}Worker, and that the backward propagation is not hindered at the worker side. Also the slice with the highest priority in the queue always gets the first preference for transmission.

We also add a producer-consumer mechanism at the receiver in the
{\named}Server in order to deal with the in-network delays. The packets received at the {\named}Server are pushed into a priority queue with the priority assigned by the {\named}Worker as the key. A server consumer thread then polls from this queue and processes the packets the same way as in a KVServer. Prioritization at the {\named}Server ensures highest priority parameters are processed first.

Apart from these modifications, we remove the explicit update notification and
pull requests from the KVServer. {\named}Server immediately broadcasts the
updated parameters to all workers once it has received all of the updates.
Since workers always issue a pull request after every push, this change does
not affect the correctness of the training algorithm. This modification was
necessary, because otherwise MXNet only issues a pull request once it has received the update notification for all the slices of a layer. Eliminating this helped to improve the bidirectional bandwidth utilization.

\section{Evaluation}\label{sec:eval}

\subsection{Methodology}\label{sec:meth}
We have evaluated the \name implementation on three image classification models: ResNet-50 \cite{resnet}, InceptionV3 \cite{incep}, VGG-19 \cite{vgg} and on a machine translation model, Sockeye \cite{sockeye}. We chose the standard MXNet KVStore implementation described in Section \ref{sec:kvs} as the baseline in all performance evaluation experiments.

Since \name implementation does not interfere with the model implementation or the training algorithm, the model convergence is not affected in any way. This means that the baseline and the \name would follow the same training curve for a given hyper-parameter set. The improvement in training performance is completely determined by the rate at which input data is processed. Therefore the primary performance comparison metric we use is the training throughput, which is the number of total training samples processed by the worker machines in one second. We measure the throughput after training the models for a few iterations until the throughput has become stable and then averaged over thousand iterations. In all the experiments, we set the number of KVServers/{\named}Servers equal to the number of worker machines.

\subsection{Summary of the experiments}
We conduct performance evaluation of \name in three different experiments. Section \ref{sec:tp-bw} shows how resilient \name is towards the bandwidth constraints in the network. We perform this experiment by training the model on a four machine cluster each equipped with one Nvidia P4000 GPU \cite{p4000} and interconnected with a 100Gbps InfiniBand network \cite{mellanox}. We measure throughput variation while artificially limiting the network interface transmission rate. This simulates more realistic networking infrastructure in modern cloud services where bandwidth is usually between 1Gbps and 25Gbps \cite{ec2info}. Section \ref{sec:netutil} shows how well \name utilizes the available bandwidth and reduces the network idle time. Finally, in Section \ref{sec:scale} we test the scalability of \name on different cluster sizes. This experiment is conducted on the AWS \cite{aws} using g3.4xlarge machine instances on a 10Gbps network.

In Section \ref{sec:dgc}, we compare the convergence accuracy for models trained using \name and compression based techniques. For this comparison study, we pick the state-of-the-art compression technique Deep Gradient Compression (DGC) \cite{dgc}. We implemented DGC on MXNet based on the details provided in the original paper and information collected from the authors. In addition to these experiments, we have also evaluated the effects of different parameter slice sizes on the training throughput in Section \ref{sec:slice-size}. We have also included additional results in the Appendix \ref{sec:evalapp} based on our reviewers' feedback.

\subsection{Bandwidth v.s. throughput}\label{sec:tp-bw}
\begin{figure*}[h!]
	\centering
	\subfigure[ResNet-50]{
		\label{fig:resnet50-tp}
		\includegraphics[width=40mm]{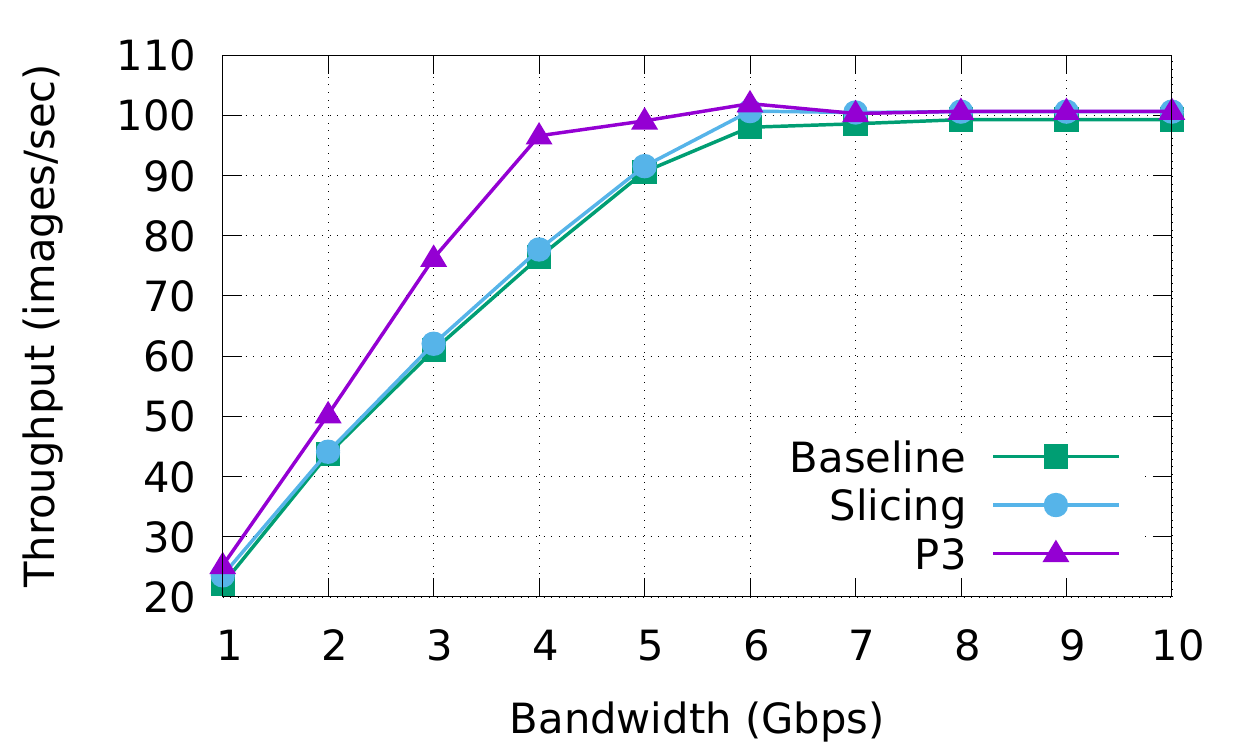}
	}
	\subfigure[InceptionV3]{
		\label{fig:incep-tp}
		\includegraphics[width=40mm]{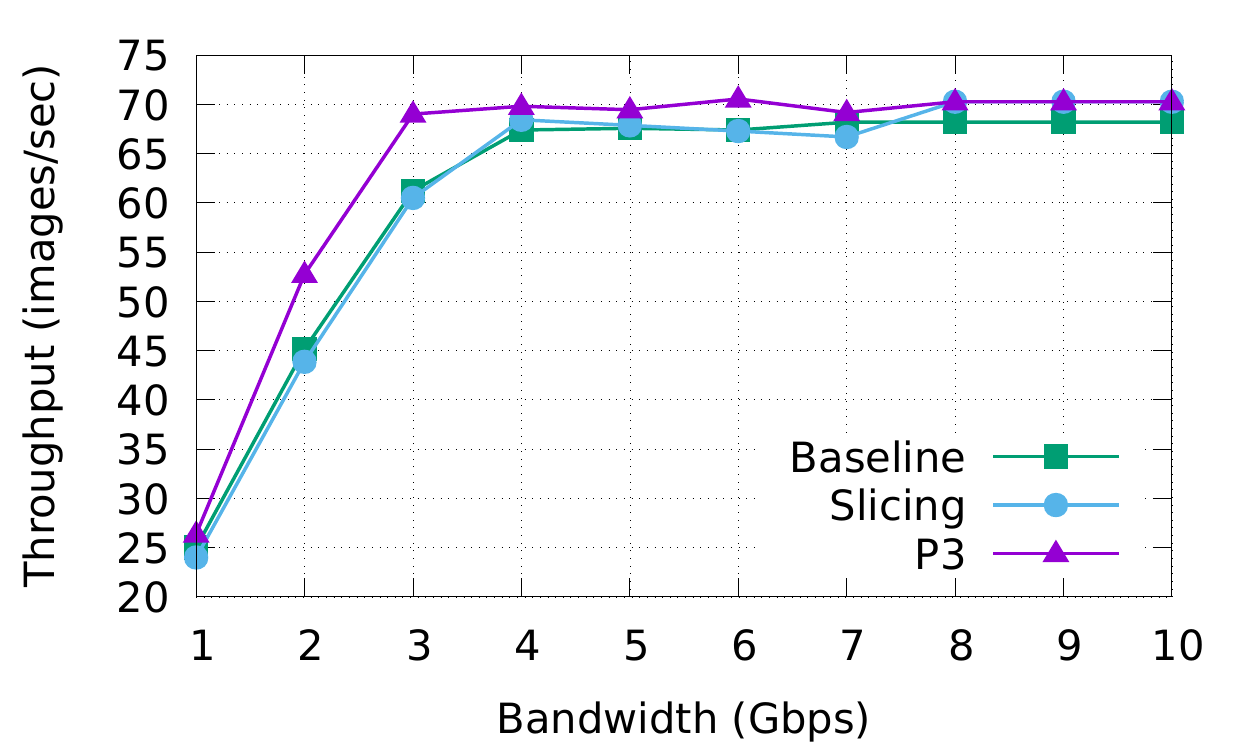}
	}
	\subfigure[VGG-19]{
		\label{fig:vgg-tp}
		\includegraphics[width=40mm]{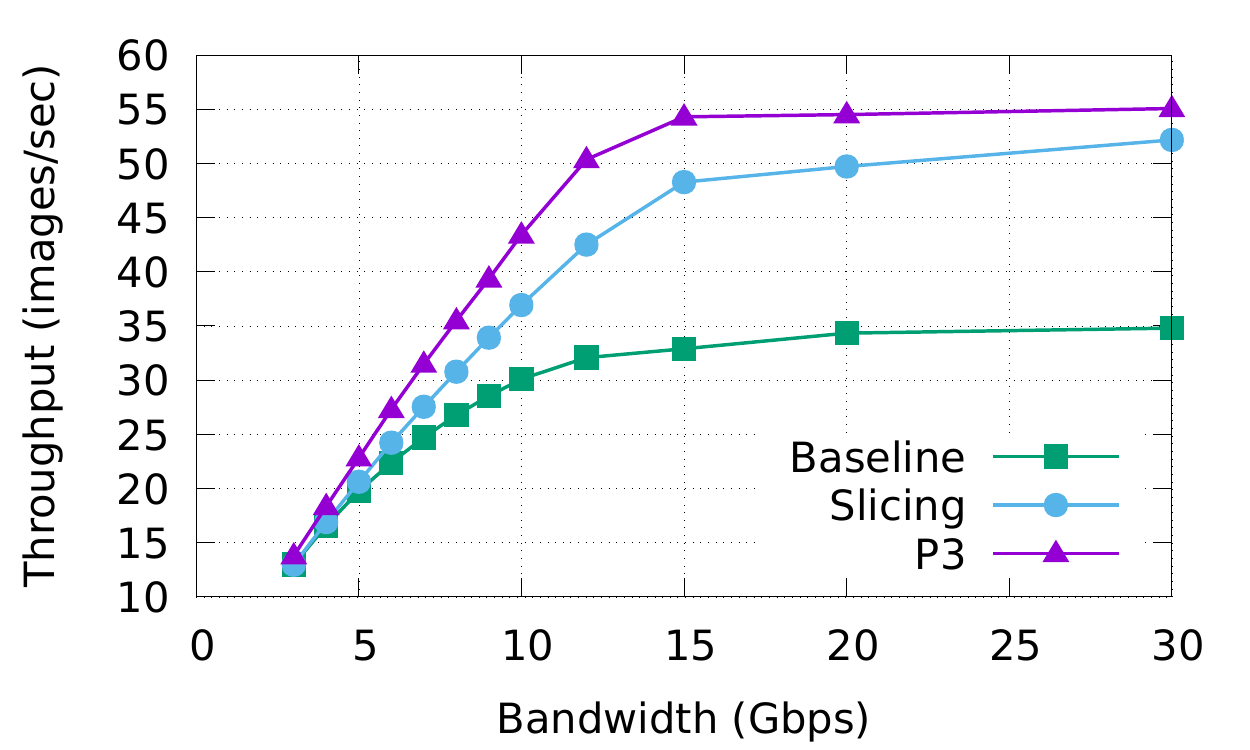}
	}
	\subfigure[Sockeye]{
		\label{fig:sockeye-tp}
		\includegraphics[width=40mm]{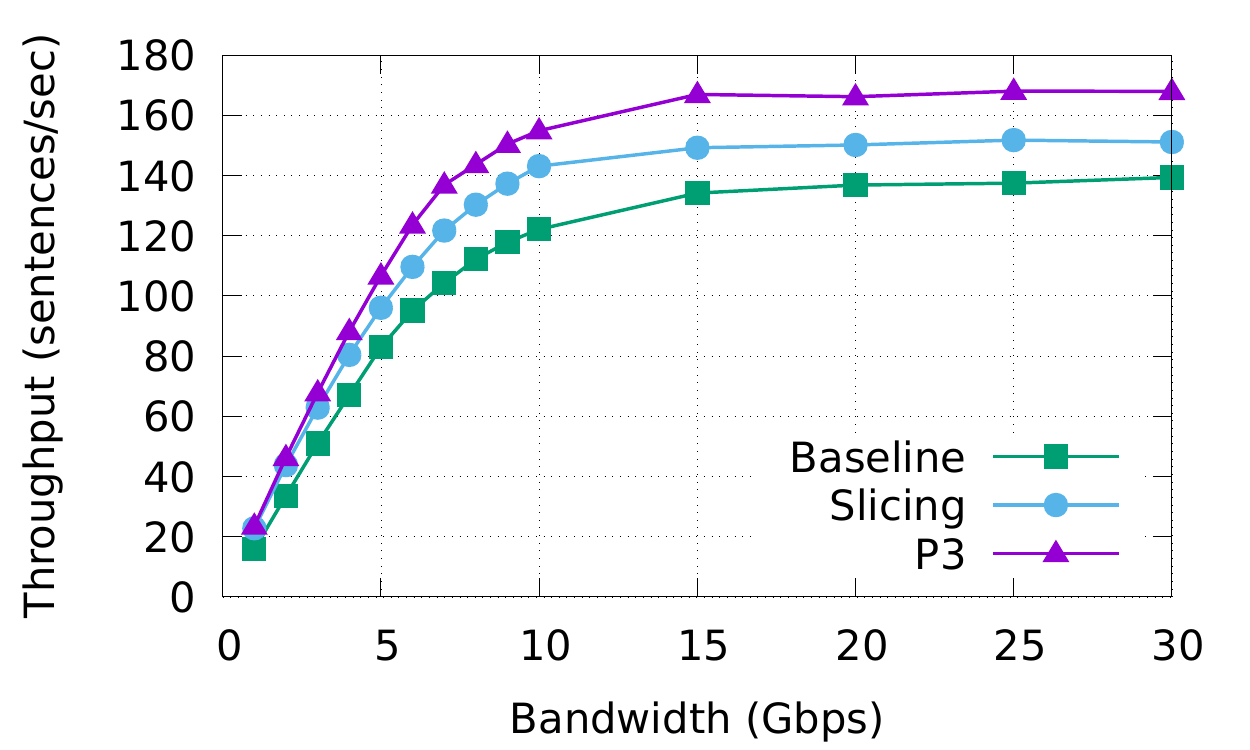}
	}
	\vspace{-5pt}
	\caption{Bandwidth v.s. Throughput}
	\label{fig:tp}
\end{figure*}

In this experiment, we measure the training throughput of ResNet-50, InceptionV3, VGG-19, and Sockeye on a tightly controlled four-machine cluster by setting different transmission rates on the network interface on all the machines using Linux's \textit{tc qdisc} utility \cite{tc}. Figure \ref{fig:tp} compares the throughput from \name against the baseline system for different network bandwidths between 1Gbps and 30Gbps. We also measured the performance benefits achieved from the parameter slicing optimization alone.

In Figure \ref{fig:resnet50-tp} and \ref{fig:incep-tp}, both baseline and \name give similar training performance when the network bandwidth is sufficiently high for scaling these models on four machines. However, the baseline throughput starts to drop in ResNet-50 below 6Gbps. At the same time, \name maintains the linear throughput until the bandwidth drops below 4Gbps. This is because \name reduces the peak bandwidth required for the model by efficiently overlapping communication with the computation. At 4Gbps, \name provides $26\%$ more throughput than the baseline. For InceptionV3, the maximum speed up obtained is $18\%$. It is interesting to note that these models do not benefit from parameter slicing, as the layer sizes are relatively small in these DNNs (see Figure \ref{fig:dist-resnet}).

Figure \ref{fig:vgg-tp} and \ref{fig:sockeye-tp} show the throughput of VGG-19 and Sockeye respectively. These models contain one or more very large layers (Figure \ref{fig:dist-vgg} and \ref{fig:dist-sockeye}), and because of the presence of these large layers, the parameter slicing optimization alone is giving considerable improvement in performance. At 30Gbps, parameter slicing can provide $49\%$ speedup on VGG-19. The speedup is further improved with \name by as much as $66\%$ at 15Gbps. Sockeye is a special case among other models. Unlike image classification models, the heaviest layer in this model is the initial layer. In Figure \ref{fig:sockeye-tp}, Sockeye performance has improved by a maximum of $38\%$ with \named.

We observe that \name always performs better than the baseline with relatively higher performance benefits when bandwidth is limited. Since \name reduces the peak network bandwidth required for the training, it is more suitable than baseline on a shared network cluster where effective bandwidth available for a single training process is much lower than the maximum capacity of the network. However, the speed-up diminishes when the network bandwidth is too low. This is because the communication time is significantly higher than computation and there is little room for improvement by overlapping communication and computation.

\subsection{Network utilization}\label{sec:netutil}
\begin{figure*}[h!]
	\centering
	\subfigure[ResNet-50 at 4Gbps]{
		\label{fig:nw-resnet-base}
		\includegraphics[width=50mm]{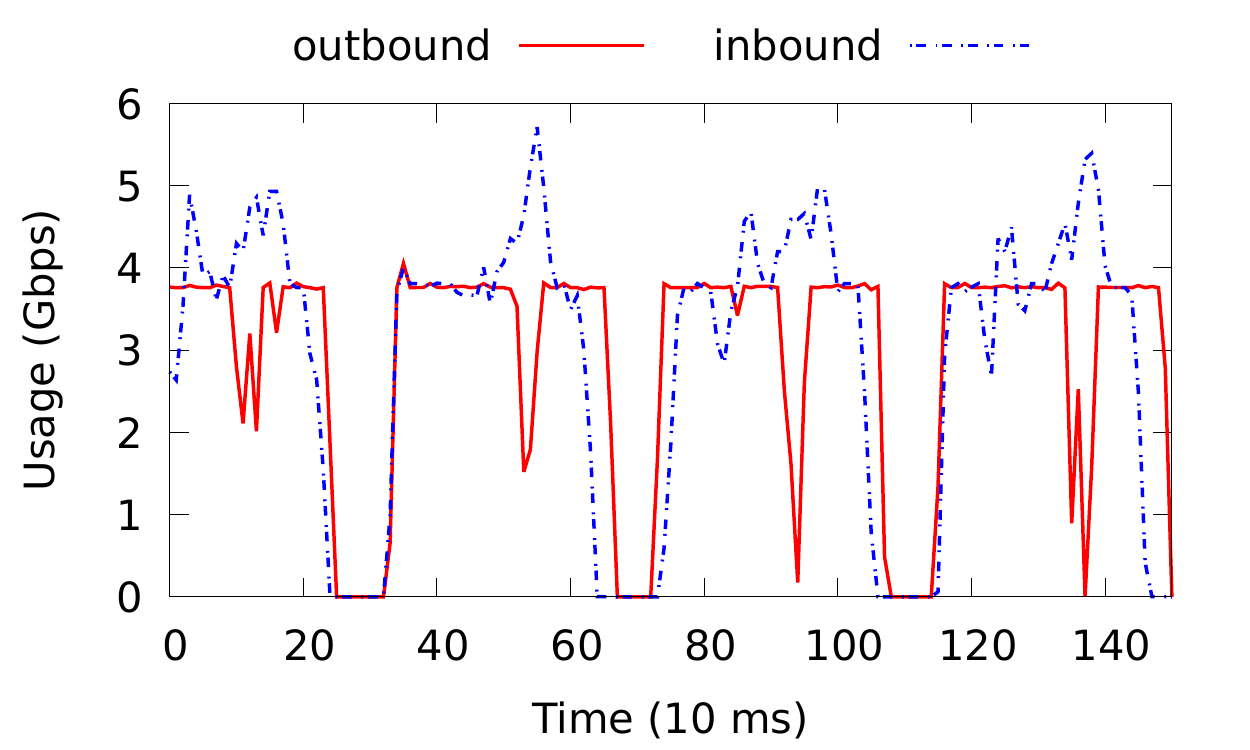}
	}
	\subfigure[VGG-19 at 15Gbps]{
		\label{fig:nw-vgg-base}
		\includegraphics[width=50mm]{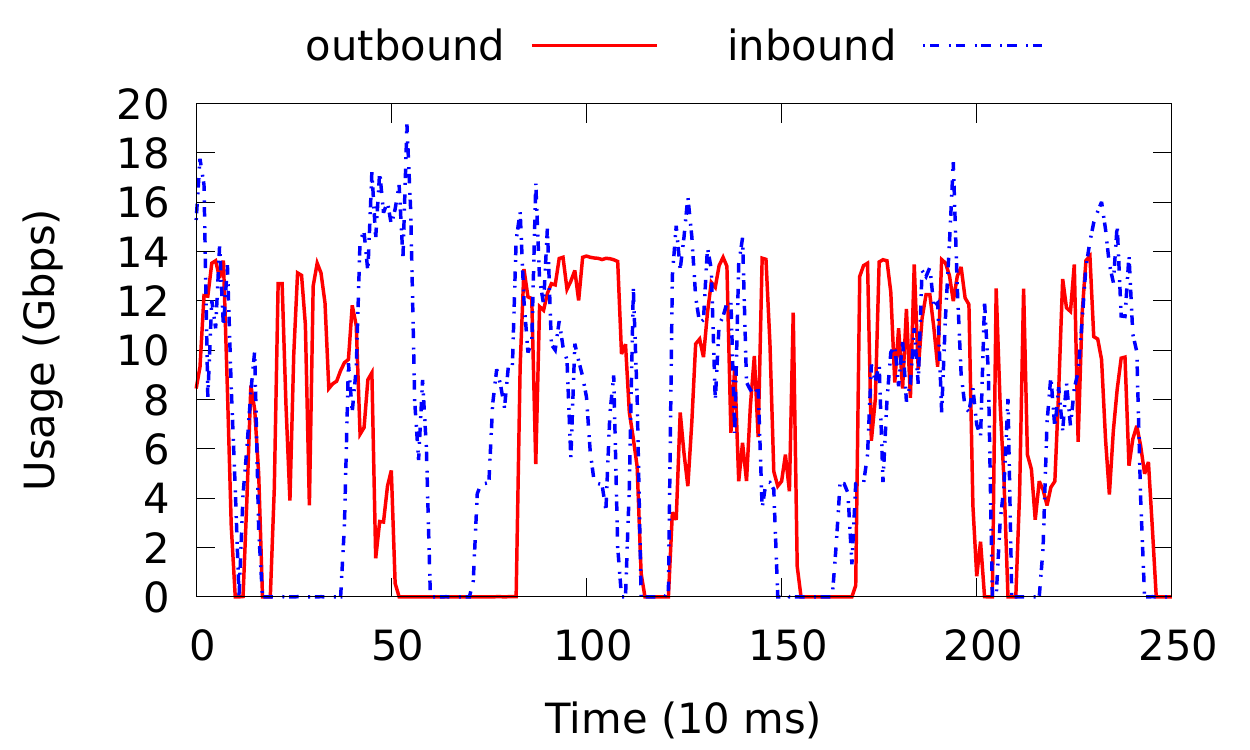}
	}
	\subfigure[Sockeye at 4Gbps]{
		\label{fig:nw-sockeye-base}
		\includegraphics[width=50mm]{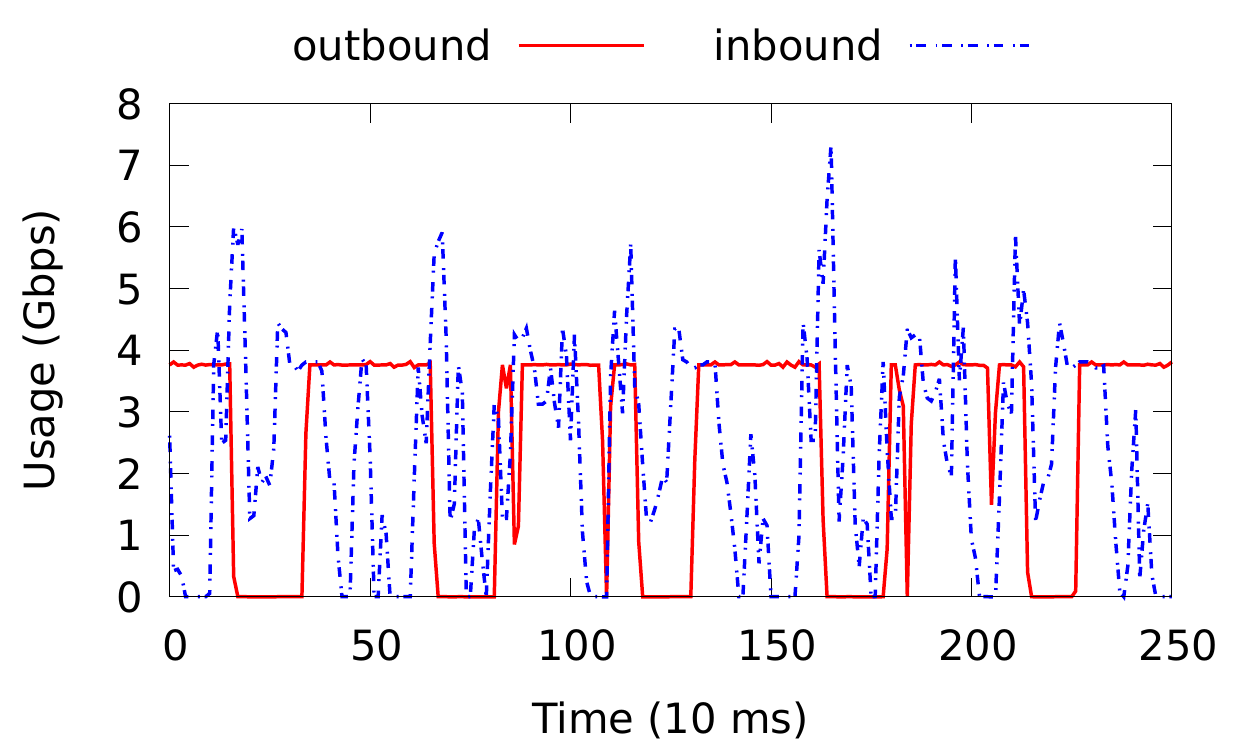}
	}
	\vspace{-5pt}
	\caption{Network utilization of the baseline system}
	\label{fig:nw-base}
\end{figure*}
\begin{figure*}[h!]
	\centering
	\subfigure[ResNet-50 at 4Gbps]{
		\label{fig:nw-resnet-prio}
		\includegraphics[width=50mm]{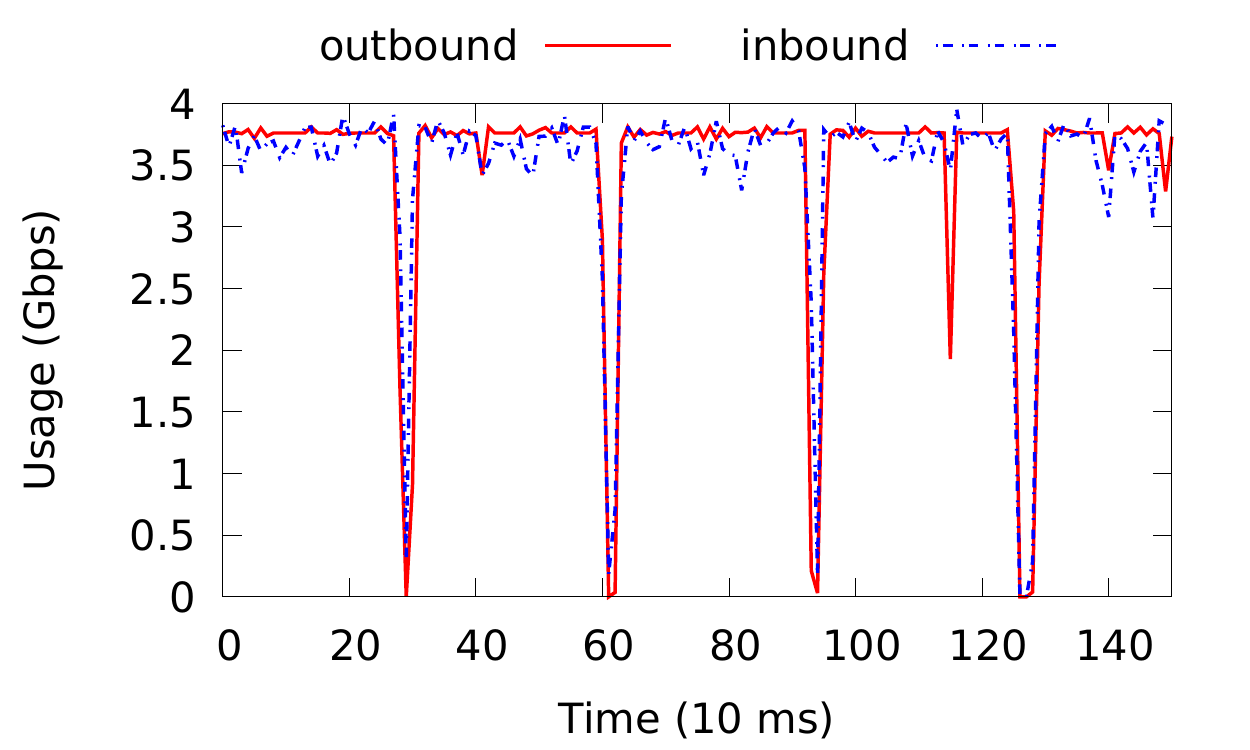}
	}
	\subfigure[VGG-19 at 15Gbps]{
		\label{fig:nw-vgg-prio}
		\includegraphics[width=50mm]{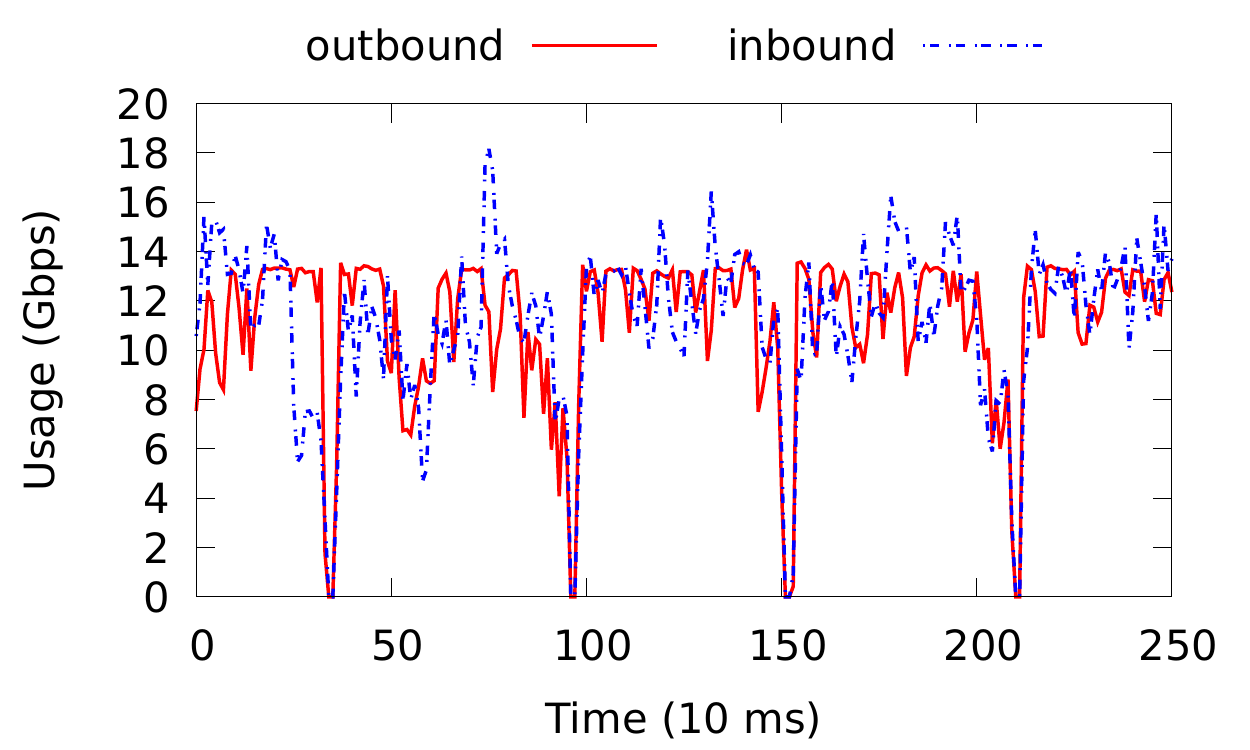}
	}
	\subfigure[Sockeye at 4Gbps]{
		\label{fig:nw-sockeye-prio}
		\includegraphics[width=50mm]{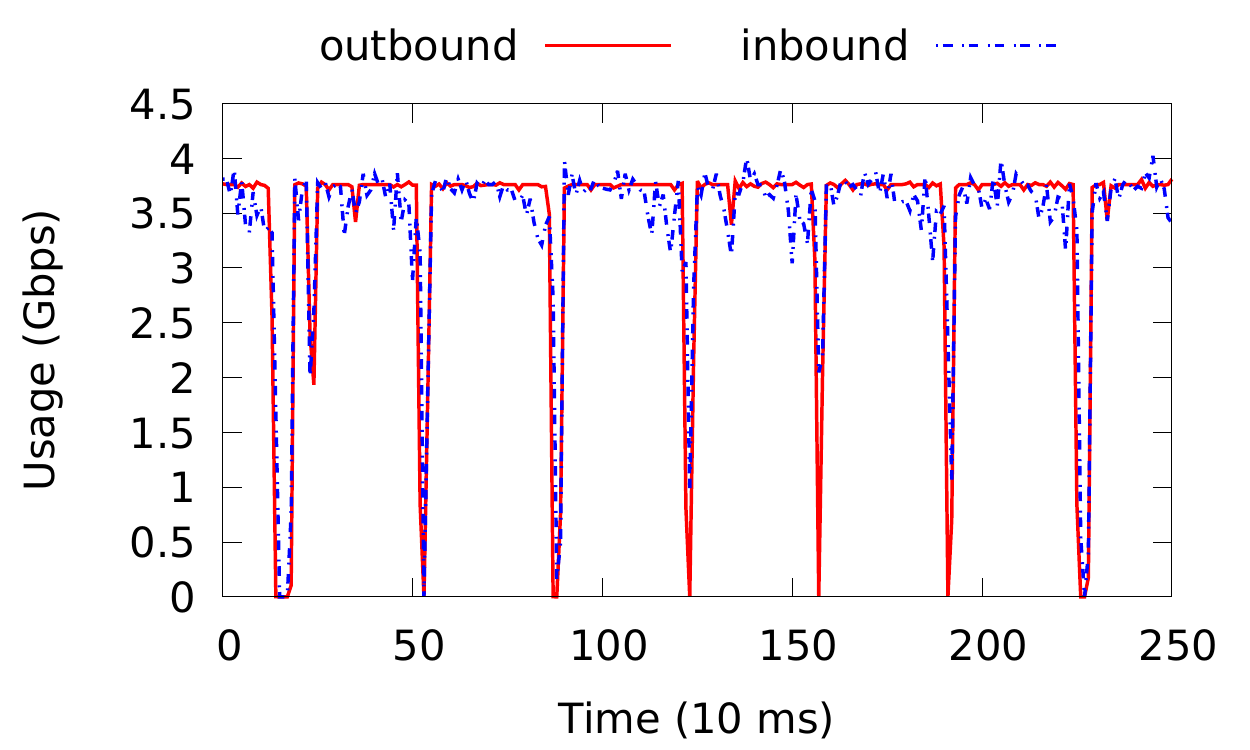}
	}
	\vspace{-5pt}
	\caption{Network utilization of \name}
	\label{fig:nw-prio}
\end{figure*}

This experiment compares the network utilization of \name with the baseline system. We conduct this experiment for ResNet-50, VGG-19, and Sockeye and measure the traffic generated from and received by one of the four worker machines. The network utilization is measured at the interface level using Linux's \textit{bwm-ng} tool \cite{bwm} with 10 millisecond precision. Figure \ref{fig:nw-base} shows the network utilization of the baseline system. The baseline implementation has bursty network traffic generated with regular peaks and crests across all models. This pattern is observed in other frameworks like TensorFlow \cite{tf} and Poseidon \cite{poseidon} as well (see Appendix \ref{sec:evalapp}). For Sockeye and VGG-19, the network idle time is extremely dominant because of the presence of heavy layers. Moreover, the inbound and outbound traffics are not overlapped as the baseline fails to fully utilize bidirectional bandwidth.

In contrast, Figure \ref{fig:nw-prio} shows the network utilization graph with \named. We observe that \name significantly improves the network utilization compared to the baseline. In Figure \ref{fig:nw-resnet-prio} and \ref{fig:nw-vgg-prio}, the network idle time is seen to be reduced with \named. Especially for Sockeye in Figure \ref{fig:nw-sockeye-prio}, \name utilizes bidirectional bandwidth more effectively than baseline system. This is one of the key reasons for the speedup observed for Sockeye despite having heavy initial layers.

\subsection{Scalability}\label{sec:scale}
\begin{figure*}[h!]
	\centering
	\subfigure[ResNet-50]{
		\label{fig:scale-resnet}
		\includegraphics[width=50mm]{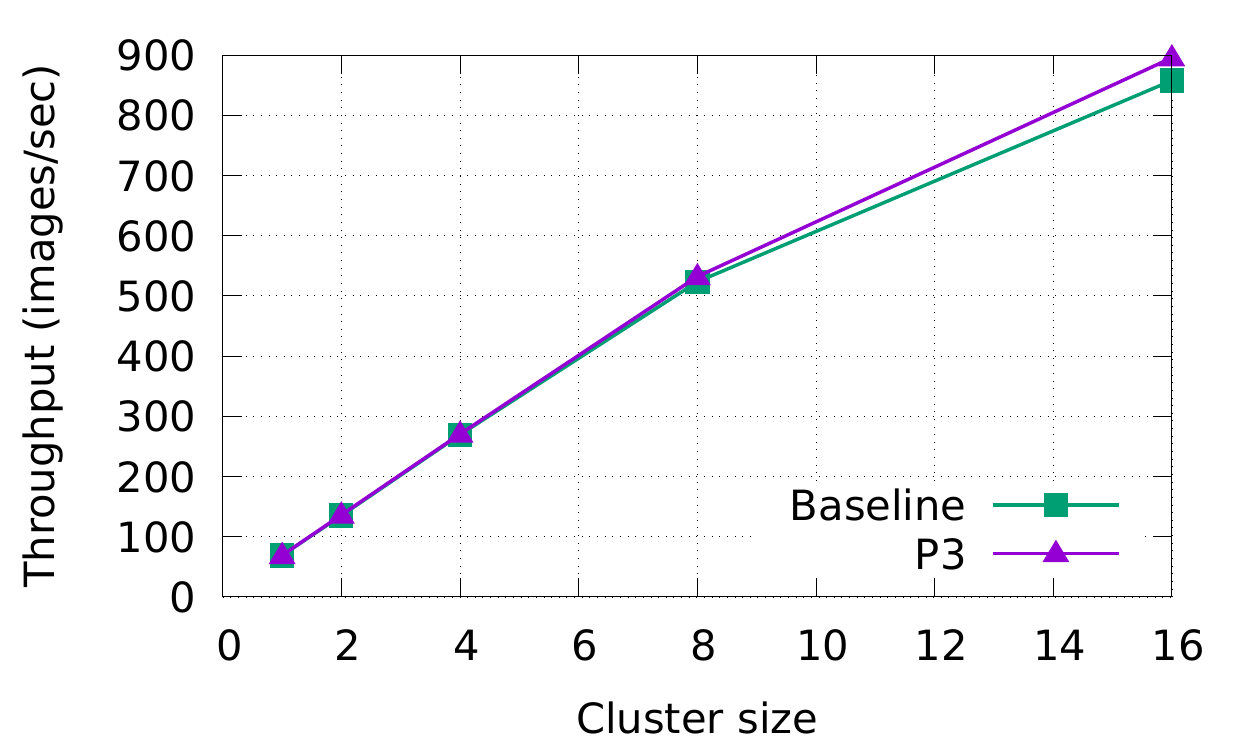}
	}
	\subfigure[VGG-19]{
		\label{fig:scale-vgg}
		\includegraphics[width=50mm]{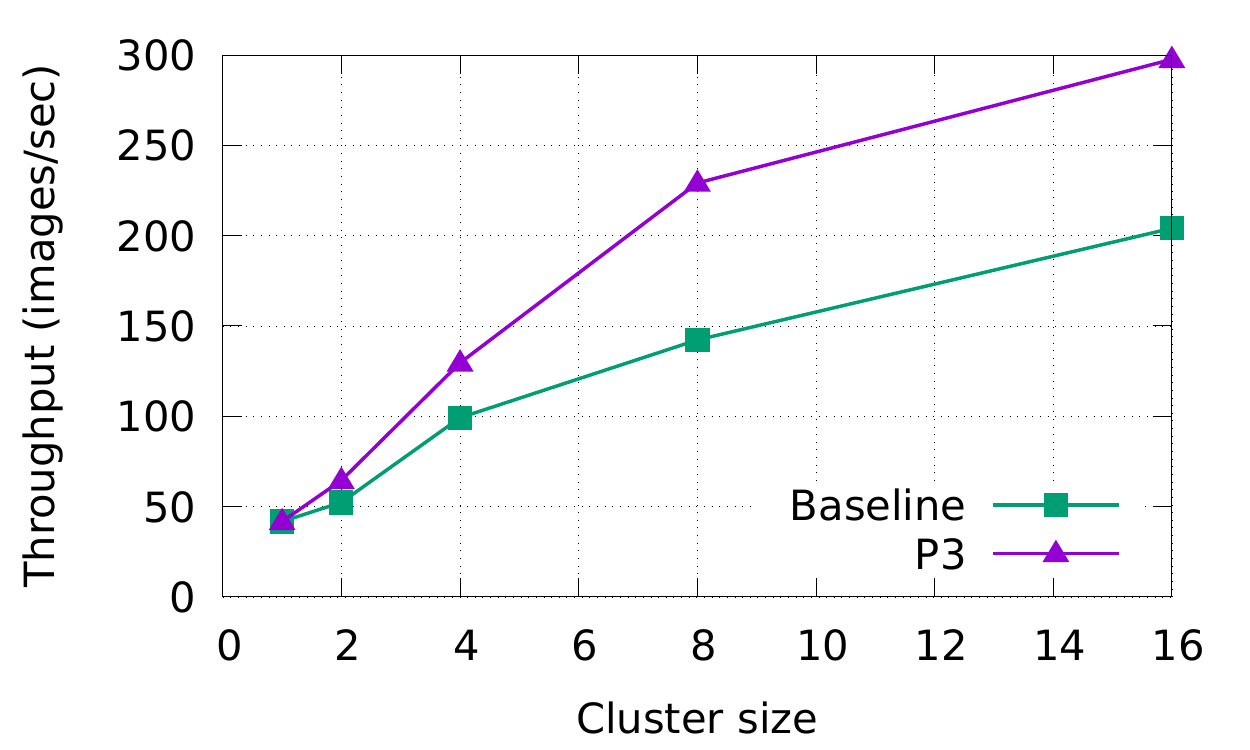}
	}
	\subfigure[Sockeye]{
		\label{fig:scale-sockeye}
		\includegraphics[width=50mm]{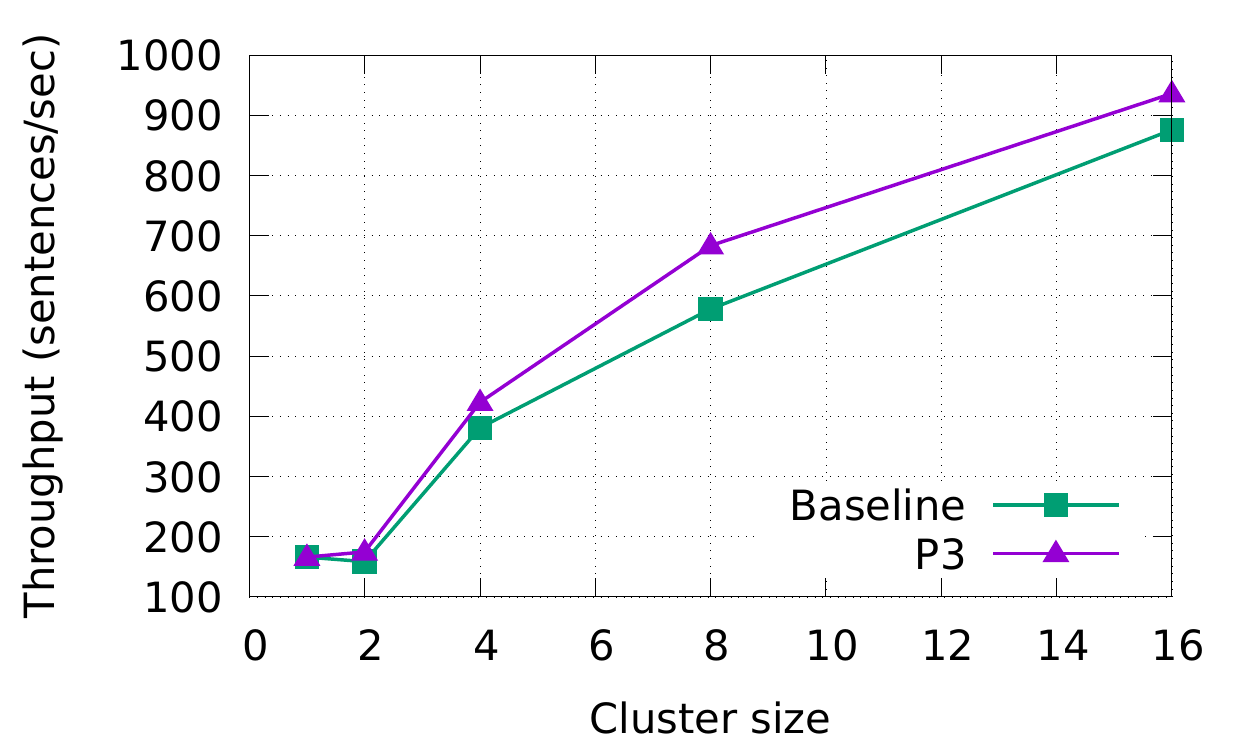}
	}
	\vspace{-5pt}
	\caption{Throughput scaling with different number of machines}
	\label{fig:scale}
\end{figure*}
We perform scalability analysis on ResNet-50, VGG-19, and Sockeye in order to show how well \name can perform on large clusters compared to the baseline system. We conduct this experiment by distributing models on AWS g3.4xlarge clusters of different sizes (2, 4, 8 and 16 machines) over a 10Gbps network.

Figure \ref{fig:scale-resnet} shows that on ResNet-50 both the baseline and the \name achieve similar performance. As shown in Section \ref{sec:tp-bw}, 10Gbps network is more than enough for linearly scaling ResNet-50. The throughput of VGG-19 has been considerably improved with \name by as much as $61\%$ on an eight machine cluster (Figure \ref{fig:scale-vgg}).

Figure \ref{fig:scale-sockeye} shows the scalability of Sockeye. LSTM-based
models like Sockeye are very hard to scale over multiple machines, because of the heavy
initial layers and difference in iteration time in worker machines due to the
variable sequence length of input data. Nevertheless, with \named, we improve throughput of Sockeye by as much as $18\%$ on an 8-node cluster.

\subsection{Training accuracy}\label{sec:dgc}
\begin{figure}[h!]
	\centering
	\includegraphics[width=55mm]{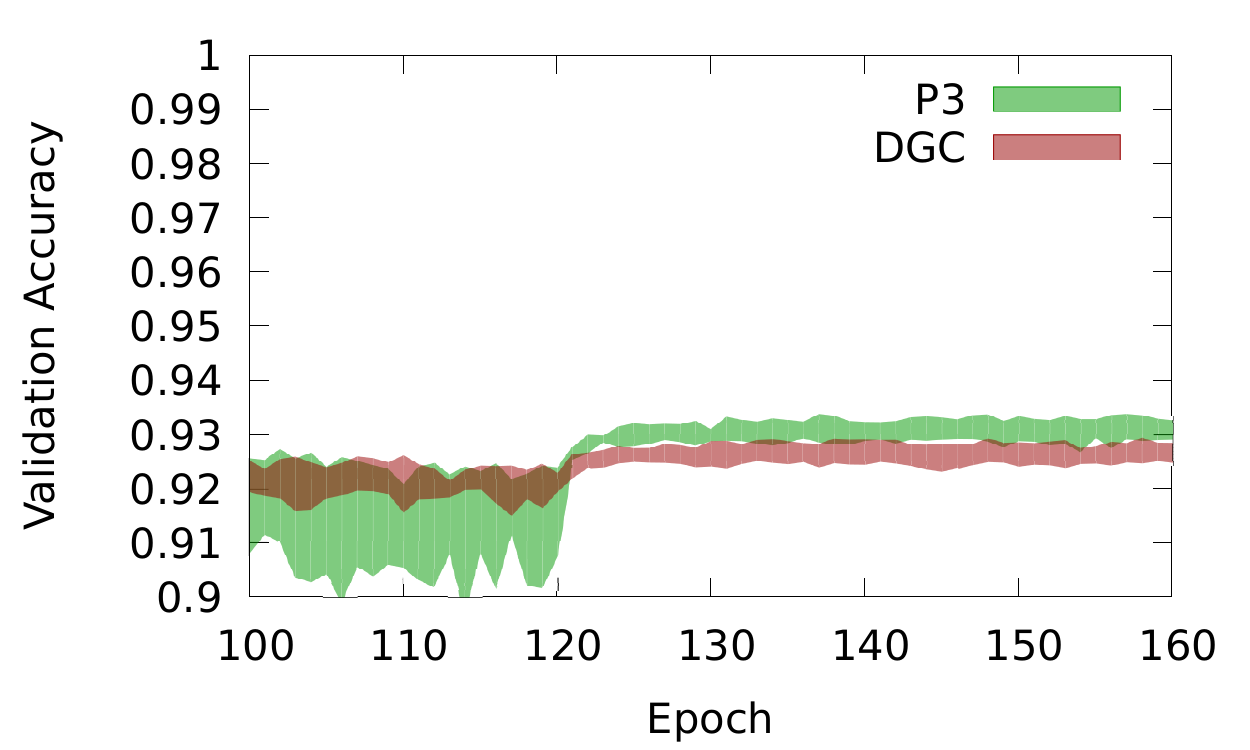}
	\vspace{-5pt}
	\caption{\name v.s. DGC}
	\label{fig:dgc}
\end{figure}

As we noted in Section \ref{sec:intro}, there are many compression techniques proposed for improving data parallel training performance. These methods can provide higher performance gains compared to \named, however, at the cost of reduction in the model quality. In this section, we compare convergence accuracy of \name with the state-of-the-art compression technique called Deep Gradient Compression (DGC) \cite{dgc}. DGC reduces the amount of data transferred by taking advantage of the sparsity in the gradient updates. The key idea is to synchronize only those parameters with top-k gradients and accumulate the rest locally. In this experiment, we use a sparsity threshold of $99.9\%$ per layer based on the configurations used in the original paper \cite{dgc}.

We trained ResNet-110 on the CIFAR-10 dataset for 160 epochs over a four-machine cluster with both \name and DGC using five different hyper-parameter settings. Figure \ref{fig:dgc} shows the validation accuracy range of \name and DGC from these experiments. The dark bands represent the gap between the worst and best accuracy on the five hyper-parameter setting. We observe that the final accuracy obtained with \name is always better than DGC. We calculate an average accuracy drop of $0.4\%$ with DGC.

Unlike compression based mechanisms (like DGC), \name always communicate the full gradients with other worker machines and does not make any modification in the original SGD algorithm. As a result, the performance benefits from \name comes without any penalty on model accuracy.

\subsection{Parameter slice size selection}\label{sec:slice-size}
\begin{figure*}[h!]
	\centering
	\subfigure[ResNet-50]{
		\label{fig:slice-resnet}
		\includegraphics[width=50mm]{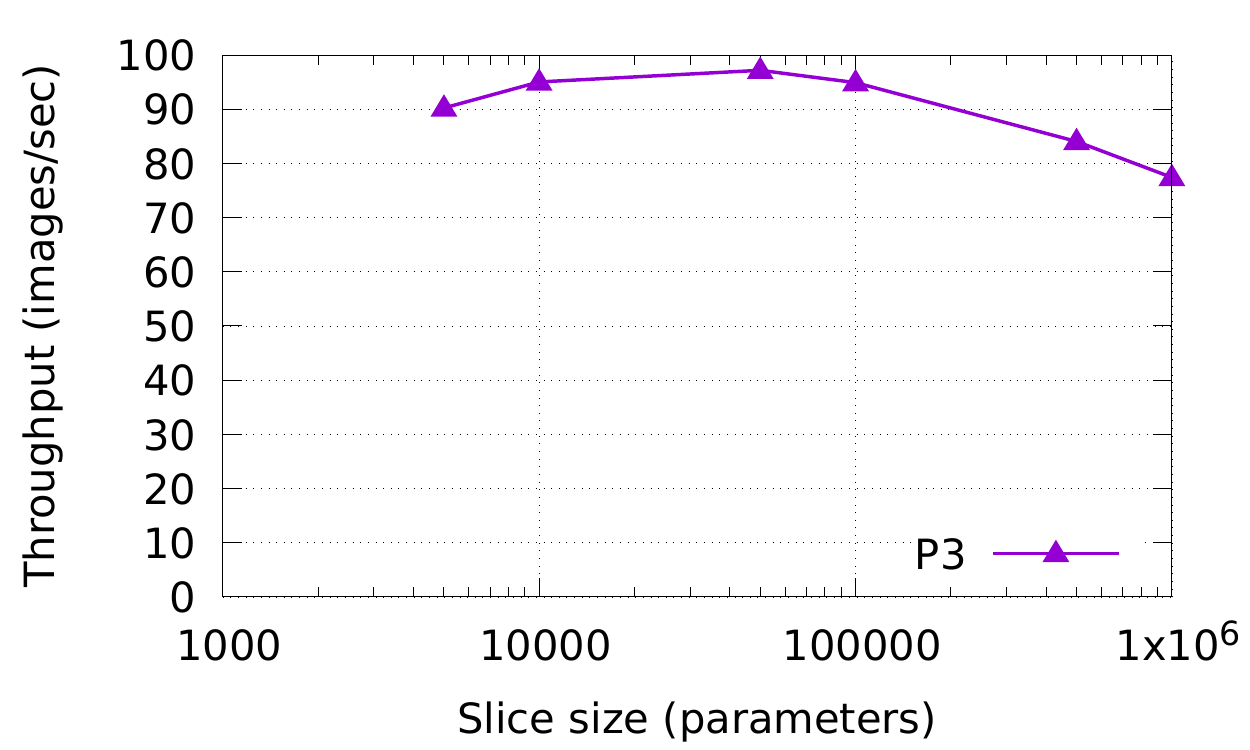}
	}
	\subfigure[VGG-19]{
		\label{fig:slice-vgg}
		\includegraphics[width=50mm]{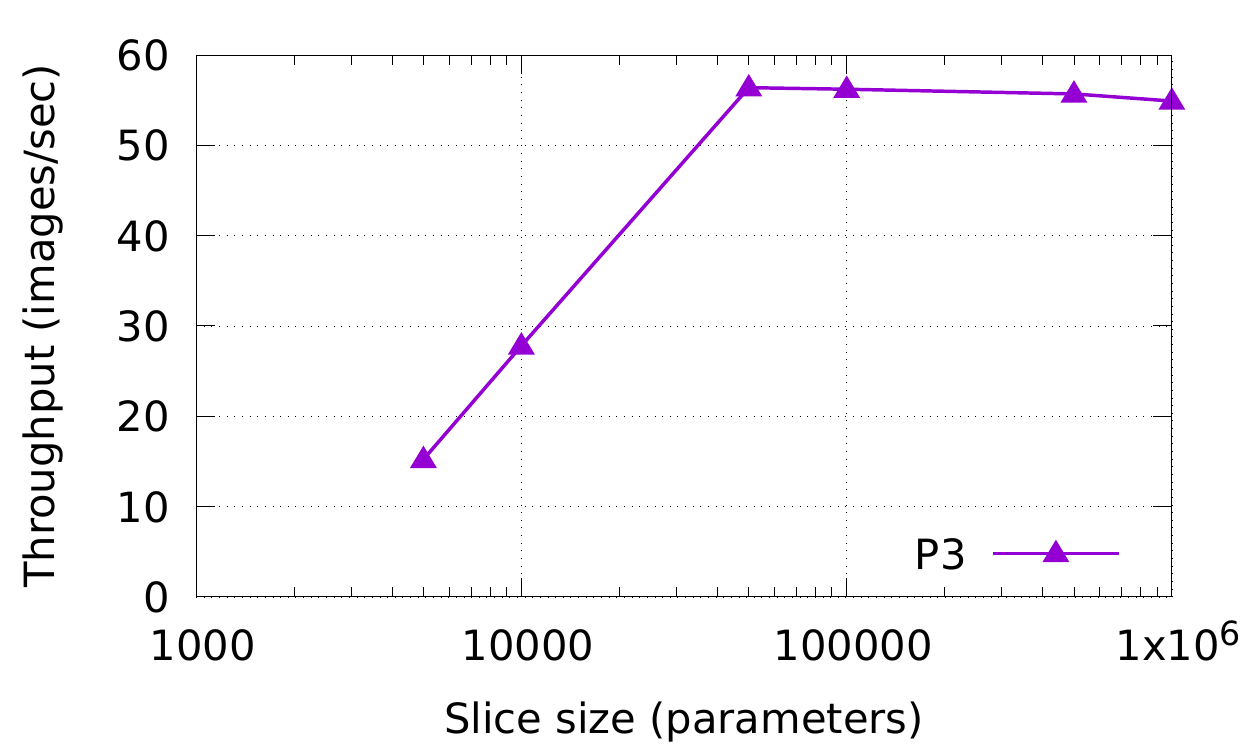}
	}
	\subfigure[Sockeye]{
		\label{fig:slice-sockeye}
		\includegraphics[width=50mm]{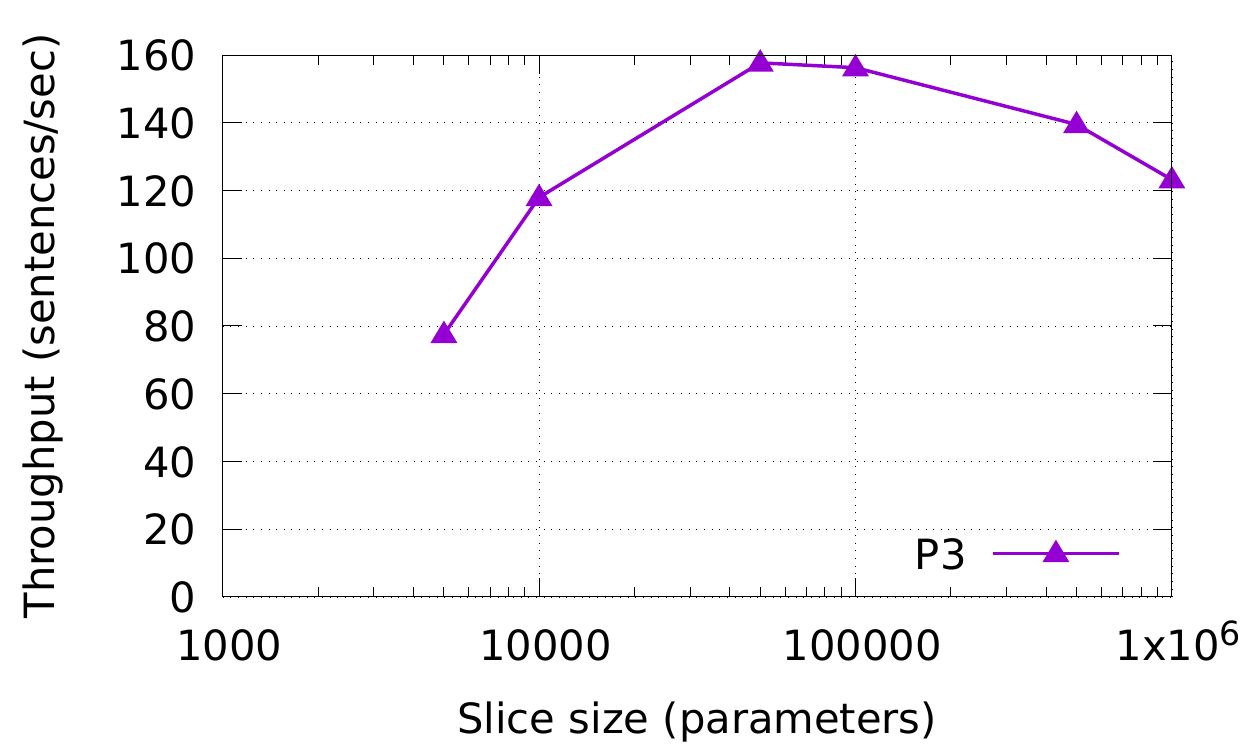}
	}
	\vspace{-5pt}
	\caption{Granularity v.s. Throughput}
	\label{fig:slice}
\end{figure*}

As we show in Section \ref{sec:prio}, a small gradient packet size can improve the network utilization and, in turn, can improve overall training throughput. In this section, we show how the size of the parameter slice affects training performance. Figure \ref{fig:slice} shows the throughput obtained for ResNet-50 and VGG-19 with \name on different parameter slice sizes.

Initially, the throughput increases as size decreases, and reaches its peak at $50,000$ parameters when it starts to drop. This happens because if the size is too small, the overhead of synchronizing packets at a very small granularity is getting too high, reducing the benefits of parameter slicing. In all our experiments, we use a maximum granularity of $50,000$ parameters per slice as it is found to be optimal empirically.

\section{Related Work}
In this paper, we describe the key limitations in the data parallel deep learning distribution techniques used in popular ML frameworks (e.g., TensorFlow and MXNet), and propose solutions to mitigate these limitations by taking advantage of domain specific characteristics of deep learning models. To the best of our knowledge, this is the first work to summarize and address these limitations.

One notable prior work which proposes domain specific optimizations for
data parallel deep learning workloads is Poseidon \cite{poseidon}. This work
introduced the idea of wait-free-back-propagation (WFBP) which hides the
communication overhead behind backpropagation by independently synchronizing
individual layers in the neural network. We build upon this idea, and show that we can overlap computation with both forward and backward propagation. We further improve this idea by using parameter slicing that utilizes network bandwidth better.

Most of the recent papers in this area try to reduce communication overhead by
sending fewer gradients. One popular method to reduce data transmission is
gradient quantization (representing the gradient values using fewer bits). For
example, 1-bit SGD \cite{1bit} represents a 32-bit floating point gradient
value in a single bit. Additionally, an error feedback is added to the SGD algorithm in order to account for the information loss that comes with the value approximation. 1-bit SGD can provide up to $10\times$ speed-up for traditional speech recognition applications. Another work called QSGD \cite{qsgd} proposes a family of compression schemes which balance the trade-off between the accuracy and gradient precision to provide good performance. Similar to QSGD, TernGrad \cite{tern} uses 3-level numerical compression to reduce data transfer in data parallel training. Both QSGD and TernGrad provide mathematical guarantees on the bounds of final model convergence accuracy. In contrast, \name always sends the full 32-bit parameter values.

Another approach is the sparse parameter synchronization. The idea is to
synchronize only a few parameters on every iteration instead of the whole
model. Gradient dropping method only synchronizes parameters which have
gradient values larger than a selected threshold. The threshold is calculated based on a
fixed compression ratio \cite{gdrop}. AdaComp \cite{adacomp} automatically
tunes the compression ratio depending on the local gradient activity and
achieves up to $200\times$ compression.

All the above techniques make the trade-off between training performance
and model accuracy because of the information loss introduced by value
approximation or stale parameter updates \cite{loss}. \name on the other hand,
does not introduce any information loss since it always sends full gradient
matrix on every iteration.

On the other hand, a more recent work called Deep Gradient Compression(DGC) \cite{dgc} offers up to $600\times$ compression and around $5\times$ speedup in low bandwidth networks while maintaining the same baseline accuracy on several DNN models. DGC use local gradient accumulation and momentum correction techniques to maintain the same accuracy. Even though the authors report no accuracy loss with DGC, there is no formal proof on the convergence guarantees cited in the paper. And as shown in Section \ref{sec:dgc}, we find it difficult to reproduce their results despite substantial effort.\footnote{This includes personal communication with the authors in order to get all their experiments correctly.} In our experiments, \name always gives better accuracy than the DGC. Based on these results, we conclude that our mechanism is a safer approach, as \name does not introduce information loss in the training algorithm and therefore there is no potential risk of accuracy loss. Moreover, our proposal is an orthogonal approach to the compression techniques and can be used on top of compression mechanisms to further improve performance.

\section{Conclusion}
In this paper, we analyze the data parallel distributed training methods used in current ML frameworks and observe that they fail to fully utilize available network bandwidth and induces high penalty on training performance under network bandwidth limitations. Based on this observation, we propose a new parameter synchronization method called \named, which improves the training performance by better utilizing the available network bandwidth. We implement \name on top of the state-of-the-art ML framework MXNet and demonstrate it to have higher resiliency towards bandwidth constraints and better scalability than the baseline MXNet implementation. With \named, we improve training throughput of ResNet-50 by as much as $25\%$, Sockeye by $38\%$ and VGG-19 by $66\%$. We also have made the source code of \name publicly available.\footnote{\href{https://github.com/anandj91/p3}{https://github.com/anandj91/p3}}

\nocite{langley00}

\bibliography{sysml}
\bibliographystyle{sysml2019}

%
%
%
%
%






\appendix
\section{Artifact Appendix}

\subsection{Abstract}

Our artifact provides the \name source code, benchmarks to regenerate the performance evaluations and scripts to install \name and run the benchmarks.

Evaluating \name require more than one machine each equipped with at least one GPU and interconnected with high bandwidth network. The artifact includes three image-classification (ResNet-50, InceptionV3 and VGG-19) and one machine translation (Sockeye) benchmarks. For image-classification benchmarks, we use ImageNet1K data set which should be prepared manually. Machine translation benchmark uses IWSLT15 data set and is included in the artifact. The benchmark script launches data parallel training process in the given list of machines and outputs the training throughput.

\subsection{Artifact check-list (meta-information)}

{\small
\begin{itemize}
  \item {\bf Algorithm:} Priority-based Parameter Propagation (\named).
  \item {\bf Program:} ResNet-50, InceptionV3, VGG-19 and Sockeye. All benchmarks are included.
  \item {\bf Compilation:} GCC 5.4 or above, CUDA 8 or above, cuDNN 6 or above.
  \item {\bf Transformations:} No transformation tools required.
  \item {\bf Binary:} Source code and scripts included to build binaries on Debian/Linux machines.
  \item {\bf Data set:} ImageNet1K data set need to be prepared manually (download from \href{http://www.image-net.org/challenges/LSVRC/2012/nonpub-downloads}{http://www.image-net.org/challenges/LSVRC/2012/nonpub-downloads}). IWSLT15 data set is included.
  \item {\bf Run-time environment: } No restriction on operating systems. Installation scripts are included for Debian based Linux operating systems (Ubuntu 16.04 recommended). Sudo access required. Software dependencies include - OpenCV, OpenBLAS, CUDA, cuDNN and python3.
  \item {\bf Hardware: } Require more than one machine (four recommended) each equipped with Nvidia GPUs and high bandwidth interconnect (at least 10Gbps).
  \item {\bf Run-time state: } There is a warm-up period for the training to stabilize. We recommend to take the measurements after skipping first 1000 iterations.
  \item {\bf Execution: } Run the benchmarks for 1000 iterations once the benchmark execution has stabilized.
  \item {\bf Metrics: } The primary metric of comparison is the average training throughput.
  \item {\bf Output: } The benchmarking scripts would output the number of input data points processed per second.
  \item {\bf Experiments: } Follow the scripts provided below.
  \item {\bf How much disk space required (approximately)?: } About 200 GB per machine should be enough for running experiments. Data set preparation might require about 500 GB.
  \item {\bf How much time is needed to prepare workflow (approximately)?: } About one hour to prepare the dataset and compile the source code.
  \item {\bf How much time is needed to complete experiments (approximately)?: } About 15 minutes per benchmark.
  \item {\bf Publicly available?: } Yes.
  \item {\bf Code licenses: } Apache License 2.0
  \item {\bf Workflow framework used?: } No
  \item {\bf Archived?: } Yes. \href{https://doi.org/10.5281/zenodo.2549852}{https://doi.org/10.5281/zenodo.2549852}
\end{itemize}

\subsection{Description}

\subsubsection{How delivered}
\name source code is publicly available on GitHub: \href{https://github.com/anandj91/p3}{https://github.com/anandj91/p3}. Latest version of \name along with benchmarks can be downloaded from \href{https://doi.org/10.5281/zenodo.2549852}{https://doi.org/10.5281/zenodo.2549852}. The unpacked artifact require less than 2 GB disk space.

\subsubsection{Hardware dependencies}
We recommend to use at least four machines each equipped with Nvidia GPUs and interconnected with at least 10Gbps network.

\subsubsection{Software dependencies}
We strongly recommend to build the code on Ubuntu 16.04 with GCC 5.4. Building the source code require CUDA, cuDNN, libopenblas-dev, libopencv-dev and python3-dev. We recommend to use CUDA 9 with cuDNN 7.

\subsubsection{Data sets}
Build and install the source code before preparing the data set. Once prepared, data set need to be copied to the exact same directory location in all the machines.

{\bf ImageNet1K:} We use ImageNet1K dataset for ResNet-50, InceptionV3 and VGG-19 which can be downloaded from \href{http://www.image-net.org/challenges/LSVRC/2012/nonpub-downloads}{http://www.image-net.org/challenges/LSVRC/2012/nonpub-downloads}. Once the data set is downloaded it need to be converted to RecordIO format. Rest of the instructions assume that all images are stored as individual image files, and images belonging to the same class are placed in the same directory. All these class directories are then in the same root directory, say \emph{imnet}.

Go to the root directory of the source code and run the following commands for preparing the data set:
\begin{lstlisting}[language=bash]
$ python tools/im2rec.py --list --recursive \
	--train-ratio 0.95 imagenet1k imnet
$ python tools/im2rec.py --resize 480 \
	--quality 95 --num-thread 28 \
	imagenet1k imnet
\end{lstlisting}

{\bf IWSLT15:} Used for Sockeye benchmark. Included in the artifact.

\subsection{Installation}
Install CUDA and cuDNN before installing \named. After that, download and unpack the artifact in all the machines at the same location. Go to the root directory and run the following command for building and installing \named:

\begin{lstlisting}[language=bash]
$ ./setup-utils/install-mxnet-ubuntu-python.sh
\end{lstlisting}

Follow the same steps for installing the baseline after switching to the \emph{baseline} branch in the git repository.

\subsection{Experiment workflow}
Before running the experiments, choose one of the machines as the master machine. Update the \emph{hosts} file at the root directory of the source code in the master machine with the IP addresses of all the participating machines. Make sure that master machine can ssh to all the machines without password.

Run the following command to execute the benchmarks:
\begin{lstlisting}[language=bash]
$ ./run_exp.sh -m (resnet|incep|vgg|sockeye)
\end{lstlisting}

Following \emph{tc qdisc} command can be used to limit the per interface peak transmission rate:

\begin{lstlisting}[language=bash]
$ sudo tc qdisc add dev <iface> root tbf rate \
<tx_rate>gbit latency 50ms burst 50kb mtu 10000
\end{lstlisting}

\subsection{Evaluation and expected result}
The script outputs the training throughput to the standard output. The average throughput can be calculated and compared with the measurements given in the corresponding figures in the Section \ref{sec:eval}. Depending on the similarity of the hardwares being used for the experiments, speedups of comparable magnitude and trends should be observed.

\subsection{Experiment customization}
The benchmark scripts can be further customized to adjust the per machine mini-batch size and \emph{hosts} file location. Full usage is given below:
\begin{lstlisting}[language=bash]
$ ./run_exp.sh [-b <mini-batch size>]
               [-h <path to hosts file>]
               -m (resnet|incep|vgg|sockeye)
\end{lstlisting}

\subsection{Methodology}
Submission, reviewing and badging methodology:
\vspace{-5pt}
\begin{itemize}
  \item \url{http://cTuning.org/ae/submission-20190109.html}
  \vspace{-3pt}
  \item \url{http://cTuning.org/ae/reviewing-20190109.html}
  \vspace{-3pt}
  \item \url{https://www.acm.org/publications/policies/artifact-review-badging}
\end{itemize}

\section{Additional Results}\label{sec:evalapp}
\subsection{Network Utilization}
In this section, we show that the limitations described in Section \ref{sec:lim} exist in other frameworks as well.

\begin{figure}[h!]
	\centering
	\includegraphics[width=60mm]{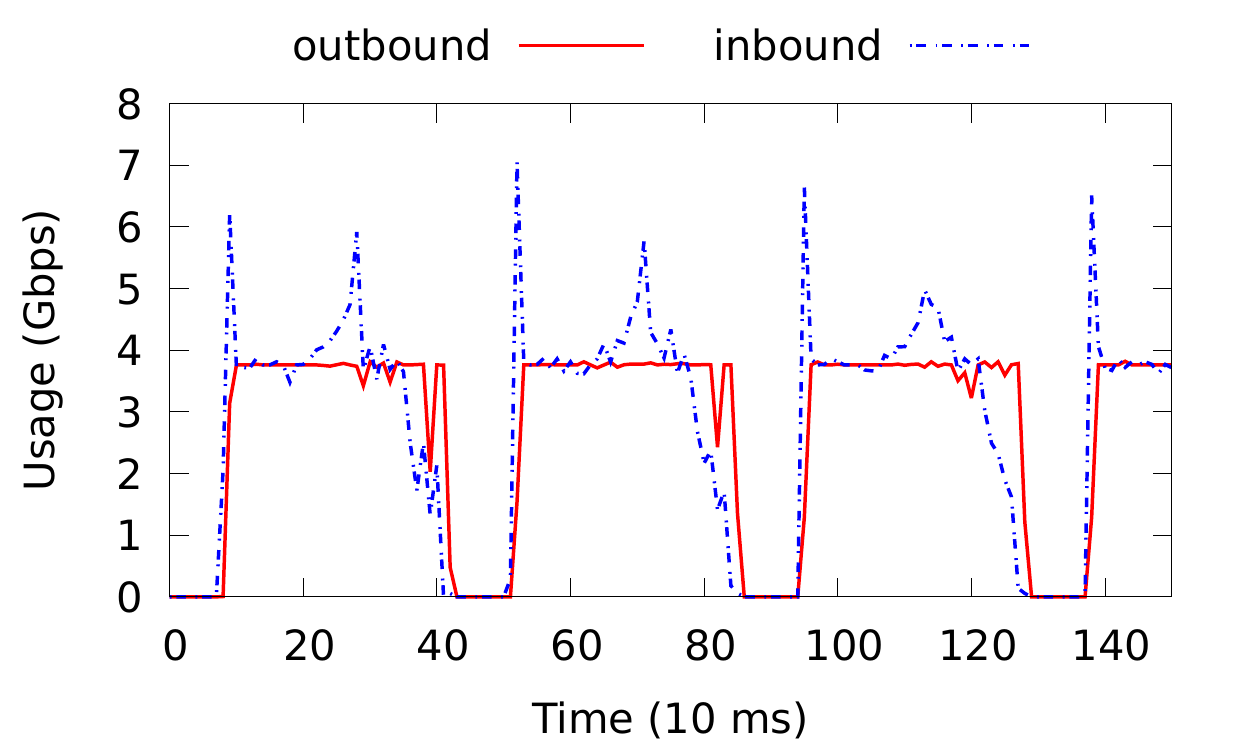}
	\caption{ResNet-50 on TensorFlow at 4Gbps}
	\label{fig:nw-tf}
\end{figure}

\begin{figure}[h!]
	\centering
	\includegraphics[width=60mm]{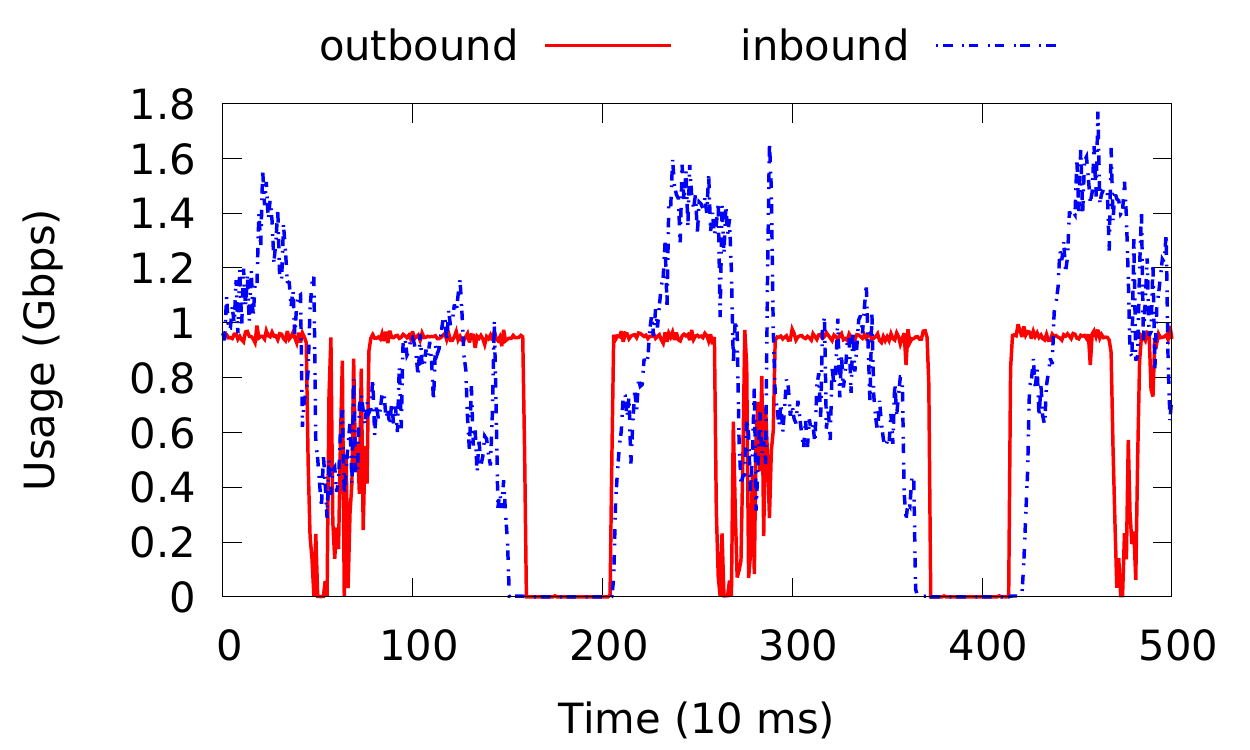}
	\caption{InceptionV3 on Poseidon at 1Gbps}
	\label{fig:nw-pos}
\end{figure}

Figures \ref{fig:nw-tf} and \ref{fig:nw-pos} show the network utilization measurements of TensorFlow and Poseidon taken while training ResNet-50 and InceptionV3 respectively on a 4-node cluster. Similar to MXNet, these frameworks also utilize network poorly even under bandwidth constraints.

\subsection{Asynchronous SGD}
ASGD algorithm does not perform synchronous update at the parameter server which means each worker machine is only blocked by its on parameter updates in an iteration as opposed to waiting for all the participating workers to finish. ASGD algorithm runs faster than synchronized SGD, however, at the cost of reduced convergence rate because of the stale parameter updates.

\begin{figure}[h!]
	\centering
	\includegraphics[width=60mm]{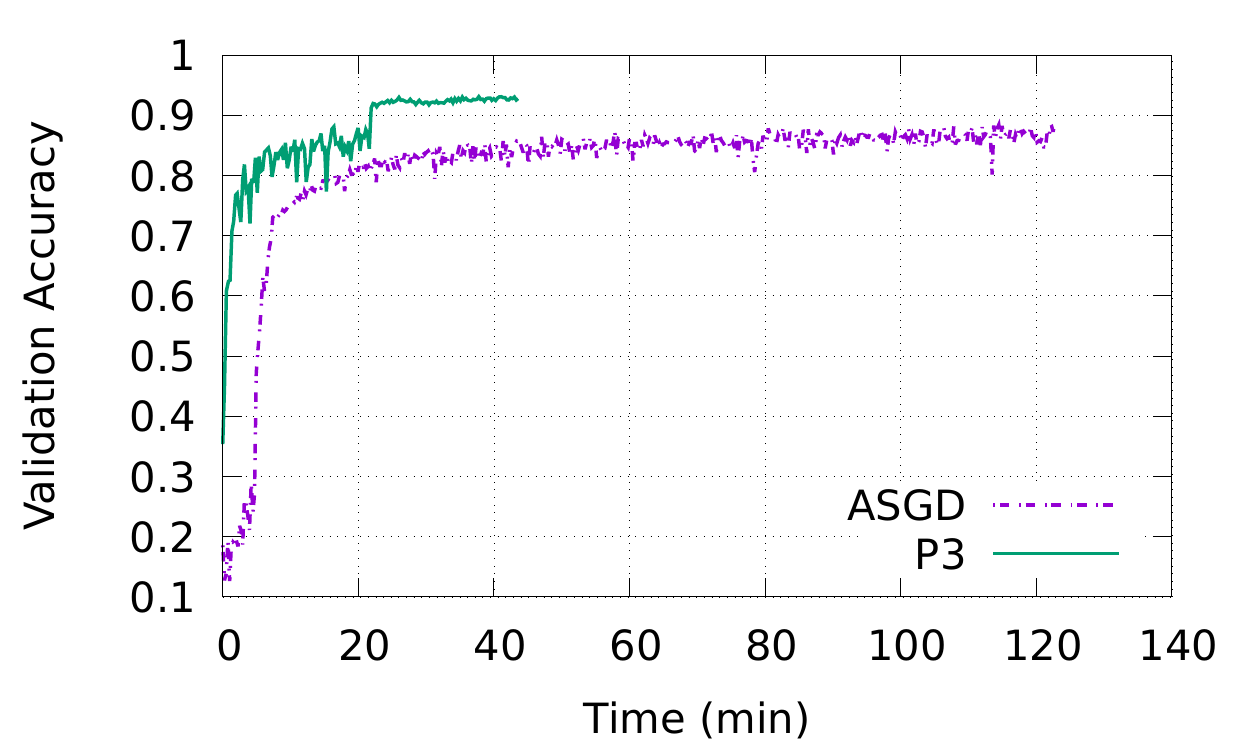}
	\caption{ASGD v.s. P3}
	\label{fig:asgd}
\end{figure}

We measured the accuracy of ResNet-110 on CIFAR-10 on a 4-machine cluster and 1Gbps network with both \name and ASGD. \name reaches a final top-1 accuracy of $93\%$ whereas for ASGD, it is only $88\%$. Additionally, \name is able to achieve $80\%$ accuracy roughly $6\times$ faster than ASGD.


\end{document}